\documentclass[aps,prd,amsfonts,amsmath,nofootinbib,tightenlines,preprint,showpacs,longbibliography,superscriptaddress]{revtex4-2}
\pdfoutput=1

\usepackage{graphicx}
    \graphicspath{ {./figures/} }
\usepackage{amsmath}
\usepackage{amssymb}
\usepackage{mathtools}
\usepackage{array}
\usepackage{hyperref}
\usepackage[caption=false]{subfig}

\def\be{\begin{equation}}
\def\ee{\end{equation}}

\def\sign{\mathop{\rm sign}}
\def\Re{\mathop{\mathrm{Re}}}
\def\Im{\mathop{\mathrm{Im}}}
\def\ohat{\hat{\mathbf{\Omega}}}
\def\bA{\mathbf{A}}
\def\bB{\mathbf{B}}
\def\bI{\mathbf{I}}
\def\bJ{\mathbf{J}}
\def\Ci{\mathop{\mathrm{Ci}}}
\def\Dai{\Delta\mathbf{a}_i}
\def\Dbj{\Delta\mathbf{b}_j}

\begin{document}

\title{Numerical gravitational backreaction on cosmic string loops from simulation}

\author{Jeremy M. Wachter}
\email{wachterj@wit.edu}
\affiliation{School of Sciences \& Humanities,\\Wentworth Institute of Technology, Boston, MA 02155, USA}

\author{Ken D. Olum}
\email{kdo@cosmos.phy.tufts.edu}
\affiliation{Institute of Cosmology, Department of Physics and Astronomy, Tufts University, Medford, MA 02155, USA}

\author{Jose J. Blanco-Pillado}
\email{josejuan.blanco@ehu.es}
\affiliation{IKERBASQUE, Basque Foundation for Science, 48011, Bilbao, Spain}
\affiliation{Department of Physics, University of the Basque Country UPV/EHU, 48080, Bilbao, Spain}
\affiliation{EHU Quantum Center, University of the Basque Country UPV/EHU, Bilbao, Spain}

\author{Vishnu R. Gade}
\email{vishnuvrgade@gmail.com}
\affiliation{Peddie School, Hightstown, NJ 08520, USA}
\affiliation{University of Illinois Urbana-Champaign, Champaign, IL 61820}
\affiliation{Institute of Cosmology, Department of Physics and Astronomy, Tufts University, Medford, MA 02155, USA}

\author{Kirthivarsha Sivakumar}
\email{kirthis2024@gmail.com}
\affiliation{Acton-Boxborough Regional High School, Acton, MA 01720, USA}
\affiliation{Institute of Cosmology, Department of Physics and Astronomy, Tufts University, Medford, MA 02155, USA}

\begin{abstract}

We performed computational gravitational backreaction on cosmic string loops taken from a network simulation. The principal effect of backreaction is to smooth out small-scale structure on loops, which we demonstrate by various measures including the average loop power spectrum and the distribution of kink angles on the loops. Backreaction does lead to self-intersections in most cases, but these are typically small. An important effect discussed in prior work is the rounding off of kinks to form cusps, but we find that the cusps produced by that process are very weak and do not significantly contribute to the total gravitational-wave radiation of the loop. We comment briefly on extrapolating our results to loops as they would be found in nature. 

\end{abstract}

\maketitle
\newpage

\section{Introduction}\label{sec:intro}

Cosmic strings are one-dimensional topological defects formed in the early universe by a spontaneously-broken symmetry~\cite{Kibble:1976} with a non-simply-connected vacuum manifold ($\mathcal{M}$): $\pi_1(\mathcal{M})\neq I$. (See \cite{Vilenkin:2000jqa} for a review.) They are a generic prediction of grand unified models of particle physics~\cite{Jeannerot:2003qv}, and can also arise from the collision of D-branes in a string-theoretic model~\cite{Sarangi:2002yt,Dvali:2003zj}. The symmetry-breaking endows the strings with a mass; depending on the details of the breaking, they might also carry currents~\cite{WITTEN1985557} or be part of a hybrid network with other defects~\cite{PhysRevLett.48.1867,PhysRevD.26.435,VILENKIN1982240}. In addition, strings are called \emph{global} or \emph{local} (a.k.a., \emph{gauge}) based on the kind of symmetry which is broken. Regardless of these details, it is generally accepted that the symmetry breaking process with $\pi_1(\mathcal{M})\neq I$ produces a network of strings which fills the universe. While there has not yet been a detection of cosmic strings, predictions of detectable signals (and constraints based on non-observations) rely principally on the characteristics of this network and the loops within it.

The detection of gravitational waves is one of the most promising channels for finding cosmic strings, particularly now that we are in the era of gravitational-wave astronomy, and many gravitational-wave observatories are searching for strings~\cite{LIGOScientific:2021nrg,KAGRA:2021tnv,NANOGrav:2023hvm,Xu:2023wog,EPTA:2023hof,NANOGrav:2023gor,Antoniadis:2023rey,Reardon:2023gzh}. In addition to sourcing gravitational waves that we might observe, gravitational effects of the string act on the loop itself in a process termed \emph{gravitational backreaction}. This self-interaction changes the shape of the loop~\cite{PhysRevD.42.2505}, in turn changing the pattern of gravitational waves emitted. Thus, an understanding of gravitational backreaction's effects on loops is critical for making precise predictions about potentially-observable signals from cosmic strings.

However, solving this problem analytically is intractable. Even for simple models of cosmic string shapes, exact solutions are known only for a few cases~\cite{Anderson_2005,Anderson:2008wa,Anderson:2009rf,PhysRevD.95.023519,Robbins:2019qba}. In a realistic network, the typical loop's shape is complex~\cite{Blanco_Pillado_2015} and not easily described by simple mathematical functions. Any large-scale study of how loops in such a network evolve must be done computationally. This of course brings its own problems, namely computational time complexity, which require the development of specialized methods and codes~\cite{PhysRevD.95.023519,Chernoff:2018evo}.

In this paper, we present the results of numerically evolving loops taken from large network simulations, under gravitational backreaction. We focus on gauge strings which are coupled only to gravity. In Sec.~\ref{sec:strings}, we review some useful properties of strings, discuss how we represent strings in our code, and introduce the corpus of loops we study in the remainder of the work. In Sec.~\ref{sec:backreaction}, we review how the numerical evolution is done, discuss the computational effort involved, and summarize the evolution of the realistic loops we studied. In Sec.~\ref{sec:power}, we discuss how we compute the gravitational radiation power from our loops. In Sec.~\ref{sec:results}, we discuss backreaction's effects on self-intersections (Sec.~\ref{ssec:intersections}), loop power spectra (Sec.~\ref{ssec:Pn}), the formation of cusps (Sec.~\ref{ssec:cusps}), and smoothing (Sec.~\ref{ssec:kink-angles}); in addition, we discuss how to extrapolate our results to loops as they might be found in nature (Sec.~\ref{ssec:kinks}). We conclude in Sec.~\ref{sec:conc}.

We work in units where the speed of light and $\hbar$ are taken to be 1.

\section{Strings and our loop population}\label{sec:strings}

The most important parameter for a string network is the energy scale $\eta$ of its associated symmetry-breaking creation process. The linear energy density $\mu$ of a string (which is also the tension) is proportional to $\eta^2$, and the core width $\delta$ to $1/\eta$. The gravitational effects of a string are given by the dimensionless quantity $G\mu$, where $G$ is Newton's constant. Current non-observations of strings set an upper bound on $G\mu$ around $10^{-10}$~\cite{NANOGrav:2023hvm}, which (assuming particle-physics-model couplings of order 1) would give $\mu\sim 10^{18}\,\text{kg/m}$ and $\delta\sim 10^{-30}\,\text{m}$.

The symmetry-breaking process produces a network of long (horizon-spanning) strings and closed loops of string; the motion and in particular intercommutation of these long strings produces further loops, and it is this population of loops which we are interested in studying. For a loop with total energy $E$, we define its invariant length as $L=E/\mu$ (to distinguish from \emph{physical} length, which generally changes as the loop stretches and contracts over the course of its oscillation). The length scales of the loops run from the astrophysical down to the microscopic, but the small values of $\delta$ we consider indicate that $\delta\lll L$ effectively always; as a consequence, we work in Nambu-Goto dynamics, treating strings as one-dimensional objects with length and tension (linear density) only.

In reality, strings are curved and have sharp bends at special points called \emph{kinks}, which are formed in pairs every time any two strings intercommute (including self-intersections). In practice, we represent our strings formed in simulation by a piecewise-linear model under the dual justification that 1) actual string curvatures on short scales should be fairly mild and only have a minor effect on dynamics, and 2) sufficient density of linear pieces accurately reproduces the dynamics of a curved string. As a one-dimensional object moving in time, a string loop therefore sweeps out a worldsheet, which can be covered by two parameters. We choose these to be the null coordinates $u,v$. Here, \emph{null} indicates that the tangent 4-vectors generated by these parameters are null; by convention, the (unit-time) tangent vector associated to $v$ is called $A'^\gamma$, and to $u$, $B'^\gamma$. The Nambu--Goto equations of motion then yield, in a conformally flat gauge, the general solution (See for example \cite{Vilenkin:2000jqa}.)
\be\label{eqn:X}
    X^\gamma(u,v) = \frac12\left[A^\gamma(v)+B^\gamma(u)\right]\,,
\ee
with $X^0=\frac12(u+v)=t$, for any position on the string worldsheet. Because of the piecewise-linear nature of the worldsheet functions $A$ and $B$, the worldsheet appears as a mosaic of parallelograms whose edges are the pieces of $A$ and $B$. We can therefore completely describe the worldsheet with two piecewise-constant functions, $A'$ and $B'$, which we write as four lists: the values taken by $A'$ and $B'$ and the amount of parameter $v$ or $u$ for which each value applies, called $\sigma_A$ and $\sigma_B$.\footnote{In the formulae for backreaction, we very commonly find $A'$, $B'$, and the $\sigma$s and very rarely find $A$ or $B$ directly. Thus this set of four lists is more convenient for computation than two functions, $A$ and $B$, from which we could extract the null vectors and edge lengths.} $A'$ and $B'$ describe the angles of any parallelogram on the worldsheet, and $\sigma_A$ and $\sigma_B$ encode the length of each parallelogram's sides. The number of values in $A'$ and $\sigma_A$ is $N_A$, and in $B'$ and $\sigma_B$, $N_B$. We refer to these as the \emph{number of segments} in $A$ or $B$, respectively. At any generic fixed-time slice of the loop, there will be $N_A+N_B$ segments visible, and it is this number which we will refer to as the \emph{segmentation} or \emph{segment count} of the loop.

With an understanding of the important qualities of strings as well as how we represent our strings numerically, let us outline how we obtained a population of realistic loops on which we study gravitational self-interactions. We generated a population of loops following the method of Ref.~\cite{Blanco-Pillado:2011egf}. We ran in the radiation era, placing Vachaspati-Vilenkin initial conditions \cite{Vachaspati:1984dz} at conformal time\footnote{Times are given in units of the spacing of the Vachaspati-Vilenkin lattice.  Starting at $\tau=6$ gives an initial condition that has a density similar to the eventual scaling regime.} $\tau=6$ and saving non-self-intersecting loops whose creation time is between $\tau=250$ and $\tau=400$ and whose size at creation obeys $L_0/d_h>0.01$. This guarantees that the loops are well inside the scaling regime~\cite{Blanco-Pillado:2011egf}. Doing so yielded 198 loops taken from the scaling population. The upper $\tau$ bound was chosen to control the total time and computational effort of backreaction: the backreaction code's runtime goes as the number of segments cubed, and loops produced at later $\tau$ tend to have more segments. This trend is important in Sec.~\ref{ssec:kinks}, when we consider extrapolating our results to real loops.  By selecting loops based on time of formation, we can be confident that our population represents a fair statistical sample, so this extrapolation, and other reported results, form an accurate general picture. This can be understood by contrast: selecting by setting an upper length bound would yield more loops from earlier $\tau$ and thus bias our statistics towards younger loops. Time of formation is a physical property of the simulation, but the distribution of lengths is itself a function of $\tau$.

\begin{figure}
    \centering
    \includegraphics[scale=1.0]{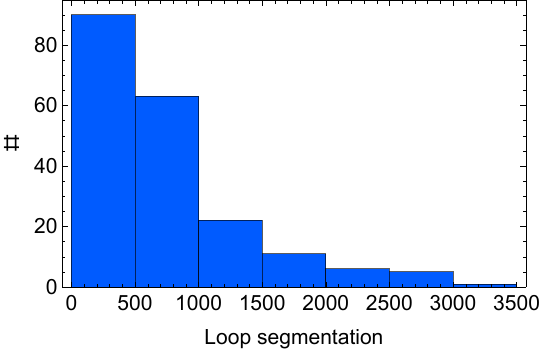}
    \caption{The distribution of segment count for the 198 loops we evolve under gravitational backreaction. All loops were produced in the radiation era in the conformal time range $\tau=250\ldots400$.}\label{fig:corpus-segmentation}
\end{figure}

The loops which make up our corpus for study range in segmentation from 187 to 3304, with a preference for low-moderate segment counts, as shown in Fig.~\ref{fig:corpus-segmentation}.
The mean segmentation is 741, and the geometric mean segmentation is 581. The particular invariant lengths of the loops are not of concern to us here---we're interested instead in how loops change with the number of oscillations, so we measure various properties of loops at fixed fractions of the initial length rather than any absolute length change. Furthermore, it is the shape, and not the scale, of the loop which determines how it changes due to gravitational self-interactions.

We have already mentioned kinks as being important structures on loops; more formally, these are any point on the string worldsheet where either $A'$ or $B'$ changes discontinuously. In the piecewise-linear representation, then, there are as many kinks as there are segments. Kinks are persistent and move at the speed of light, which the piecewise-linear representation makes easy to see: trace along the edges of the parallelograms forming a strip of the worldsheet and you will be able to track the same part of $A'$ or $B'$ around the loop in one period. In addition, loops might contain \emph{cusps} \cite{Turok:1984cn}, which are points where $A'$ and $B'$ have the same value (if we think of the three-vector parts of $A'$ and $B'$, which have unit magnitude, painted on the unit Kibble-Turok \cite{Kibble:1982cb} sphere, cusps are anywhere the two lines cross). Cusps are by nature transient---they only occur at a particular combination of $u$ and $v$---and repeat once per oscillation of the loop. In a piecewise-linear representation, where the $A'$ and $B'$ on the sphere appear as ordered sets of points, there are technically no cusps. However, if we imagine lines joining sequential points, we can identify ``crossings'' indicating where cusps might appear in an infinite-resolution simulated loop. This is the approach we use later on to discuss the effect of gravitational backreaction on the appearance of cusps.

As a final note, the loops in the simulation had center-of-mass velocities, but we boosted them to the rest frame for later analysis.

\section{Gravitational backreaction}\label{sec:backreaction}

Numerical backreaction for cosmic string loops was first implemented in~\cite{PhysRevD.42.2505}. Our backreaction code follows the same general principles as Refs.~\cite{PhysRevD.95.023519,Blanco-Pillado:2019nto}, which we summarize here.  Techniques of this kind were first used by Allen and Casper \cite{Allen:1994iq}. Our process is applied to the loops produced by the method of Sec.~\ref{sec:strings} after that network evolution code is completed.

In the absence of gravitational self-interactions, a loop which is non-self-intersecting will oscillate forever with a period of $T=L/2$, corresponding to a range $0\ldots L$ in both $u$ and $v$. In the presence of a gravitational field, however, the loop position of Eq.~(\ref{eqn:X}) is corrected by an acceleration term,
\be\label{eqn:Xuv}
    X^\gamma_{,uv} = \frac14\Gamma^\gamma_{\alpha\beta}A'^\alpha B'^\beta\,,
\ee
where $\Gamma^\gamma_{\alpha\beta}$ is the Christoffel symbol. Changes to the spatial components of the tangent vectors tell us how the shape of the string evolves, while changes to the temporal components we interpret as a loss of length, which is radiated into gravitational waves.

We do backreaction calculations only for loops in
non-self-intersecting trajectories.  (Backreaction may modify the loop
shape so that it becomes self-intersecting.  This is discussed in
Sec.~\ref{ssec:intersections}.)  We accumulate the effect of
Eq.~(\ref{eqn:Xuv}) over one oscillation, producing a change in our
tangent vectors of
\begin{subequations}\label{eqn:AB}\begin{align}
    \Delta A'(v) = 2\int^L_0 X_{,uv}(u,v)\,\text{d}u\,,\\
    \Delta B'(u) = 2\int^L_0 X_{,uv}(u,v)\,\text{d}v\,.
\end{align}\end{subequations}
The technique allows us to separate out gauge effects.  After one
oscillation (during which we do not allow gravitational effects to
perturb the string motion), the spacetime returns to its previous
metric.  Thus changes to the tangent vectors in Eq.~(\ref{eqn:AB}) are
real effects, not gauge artifacts.

To find the shape of the string loop after many oscillations, we cast
the backreaction process as a set of ordinary differential equations.
We define a new independent variable $s$ via
\be\label{eqn:s}
\text{d}s = \frac{2}{L}G\mu \text{d}t
\ee
where $L$ is the length of the loop, which decreases over time.  One
can think of $s$ as $N_\text{osc} G\mu$: the number of oscillations
that loop has undergone times $G\mu$.  Thus one application of
Eq.~(\ref{eqn:AB}) advances $s$ by $G\mu$.  Technically $s$ is a
discrete variable, but since $G\mu\ll 1$ we can treat it as
continuous.

In $X_{,uv}$ there is an overall factor $G\mu$ from the Christoffel
symbol/metric perturbation of the string.  This cancels the factor
$G\mu$ in $s$, so that the evolution of a string loop in coordinate
$s$ is independent of $G\mu$.  Furthermore this evolution is
independent of rescaling the string.  If we magnify the string by
factor $\lambda$, the metric at corresponding points is unchanged and
the Christoffel symbol decreases by $\lambda$.  In
Eq.~(\ref{eqn:AB}), the integral is for a parameter interval $\lambda$
times larger, so the change in one oscillation, and thus in interval
$G\mu$ of $s$, is unchanged.

We represent the string position using Eq.~(\ref{eqn:X}), but it is
redundant to represent $A(v)$ (and analogously $B(u)$) in terms of a
sequence of unit null tangent vectors $A'_i$ and the amount of
parameter $\sigma^A_i$ spent going in each direction.  Each $A'_i$ has
4 components, but there are two constraints: that it have unit time
component and that it be null, so there are only two degrees of
freedom.  So, for of the purpose of the differential equation, we
define $\Dai$ to be the spatial part of $\sigma^A_i A'_i$, which
is also the vector lying along the edge of each parallelogram made
from $A'$.  This is a general 3-vector without any constraint.

The function $A(v)$ is encoded as a sequence of $\Dai$.  To get
back from these vectors to the function, we construct a tangent
4-vector $A'_i$, whose time component is 1 and whose spatial component
is the unit vector in the direction of $\Dai$.  We let $\sigma^A_i
= |\Dai|$.  The function $A'(v)$ is then a sequence of $A'_i$,
each being in effect for a range of length $\sigma^A_i$ in $v$, and
$A(v)$ is its integral.  Everything here applies equally to $B$.
The combination of $A$ and $B$ in Eq.~(\ref{eqn:X}) then automatically obeys
the conformal gauge conditions.

When we represent the string using $A'_i$ and $\sigma^A_i$, there is an extra step in applying Eq.~(\ref{eqn:AB}), described in section II.B of \cite{Blanco-Pillado:2019nto}.  When we add $\Delta A'_i$ to $A'_i$, we must adjust the resulting spatial length and time component to be unit.  The new $\bA'_i$ is then
\be
\frac{\bA'_i+\Delta \bA'_i}{1+\Delta A'^t}
\ee
and the new $\sigma^A_i$ is $(1+\Delta A'^t)\sigma^A_i$.  Multiplying
these shows that the new $\Dai$ is just $\Dai+\Delta\bA'_i\sigma^A_i$.  The
same applies to $\Dbj$.

Because $G\mu\ll 1$ it is unimportant to make changes inside of a
single oscillation. So we accumulate the change as in
Eq.~(\ref{eqn:AB}) for one oscillation, but then treat it as occurring
continuously over an interval $\delta s = G\mu$, and so write
\be\label{eqn:dads}
\frac{d\Dai}{ds} = \frac{\Delta\bA'_i\sigma^A_i}{G\mu}\,.
\ee
The same applies to $\Dbj$.  As the calculation described by Eq.~(\ref{eqn:AB}) is restricted to the lowest linear order in $G\mu$, the resulting approximation given by Eq.~(\ref{eqn:dads}) may possibly lead to an error of the order $(G\mu)^2$ at a single step.  Even if this were to accumulate
linearly, the maximum error for the the entire simulation would be
$O(G\mu)$, which is negligible.  The fact that $G\mu\ll 1$ also allows
us to work in linearized gravity.

There are $N_A$ of the $\Dai$ and $N_B$ of the $\Dbj$,
each with 3 components.  Thus there are $3(N_A+N_B)$ coupled
differential equations.  We solve them using the DOP853 \cite{DOP853}
technique, an eighth-order explicit Runge-Kutta method originally due
to Dormand \& Prince.  This automatically adjusts the step size in $s$
to produce a specified accuracy. We then record the shape of the loop
every time that $s$ increases by $10^{-5}$.

Backreaction may accelerate the loop.  (Equivalently, gravitational
waves may be emitted anisotropically leading to the ``rocket
effect''.)  We would like to keep our loops in the rest frame, so we
apply small adjustments to the changes coming from Eq.~(\ref{eqn:AB})
to apply a tiny Lorentz boost (proportional to $G\mu$) to return the
loop to the rest frame.

In order for the loop to be closed, the total of the vectors $\Dai$ and of the $\Dbj$ must be the same, and the total length of
these vectors must also be the same.  Exact evolution would preserve
these constraints, but because we calculate the backreaction only in
the center of each segment, there is a small error.  So we make the
minimal change required to the results of Eq.~(\ref{eqn:AB}) so that
the loops will remain closed.  This change is applied to the various
$d\Dai/ds$ and $d\Dbj/ds$ in proportion to their length.

\subsection{Faithful representation of kink smoothing}\label{ssec:kink-smoothing}

We have made a number of other improvements to the code base first used in Ref.~\cite{Blanco-Pillado:2019nto}, such as speeding up computations of the perturbation due to each segment, parallelizing the code, and improving the numerical stability. One change particularly relevant to the results we present here has to do with how we approximate actual strings---which are smooth, except at kinks, and generically curved---by our piecewise-linear model. We wish to capture something about the ``rounding off'' or ``filling in'' of actual kinks, about which we know that 1) the scale of kink rounding grows as a cube root in time, and 2) only the part ``below'' the kink (the side with null parameter $u$ or $v$ smaller than the value of that parameter at which the kink occurs) is rounded. This is due to the self-force diverging (in an integrable way) for an observer approaching the kink from below, but remaining bounded when approaching from above. See Ref.~\cite{Blanco-Pillado:2018ael}, particularly Fig.~1 and surrounding discussion, for a detailed explanation of this one-sided rounding-off.

\begin{figure}
    \centering
    \includegraphics[scale=1.0]{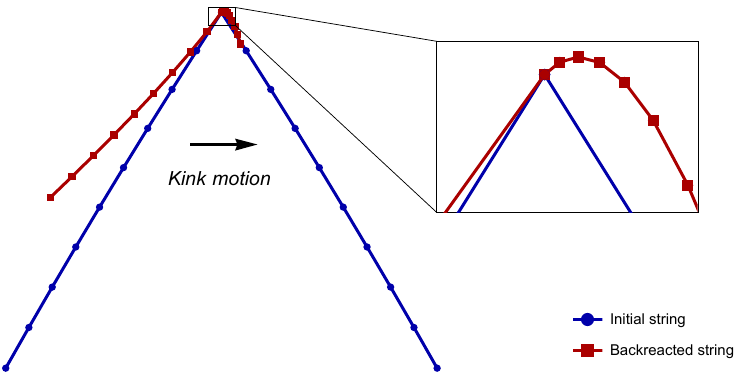}
    \caption{To faithfully represent the rounding-off of a kink by backreaction, more points are needed on one side of the kink. The blue lines represent segments of a string with a kink prior to backreaction; the red lines represent those same segments after backreaction. The segments ``below'' the kink---in the direction of kink motion---experience more length loss and bending due to backreaction, and thus all nine segments are necessary to capture the evolution, but ``above'' the kink could be faithfully represented with fewer points.}\label{fig:kink-bending}
\end{figure}

To capture this effect, we want sufficient resolution in our piecewise-linear model before large kinks, as illustrated in Fig.~\ref{fig:kink-bending}.
Adding a higher density of segments is only needed on one side of the kink; ``above'' the kink, the string has changed so little that perhaps two segments can fairly well represent the range of string shown there.

Sufficient resolution is achieved by splitting segments which are too large, by some metric, into smaller segments, in a sort of adaptive string refinement method. Consider a segment of length $d$, which is followed by an angle $\theta$, on a string of length $L$. We want to use longer segments over mostly-straight sections of string, and shorter segments in highly-curved sections of string (as we anticipate needing just below kinks). Thus, we say there is some maximum angle $\theta_\text{max} = cL/d$ a segment of length $d$ can represent, where $c$ is a constant of proportionality. If $\theta\leq\theta_\text{max}$, no action is needed. But otherwise, we must break a long segment into $M$ shorter segments, each of length $d/M$, and each of which will be responsible for representing an angle $\theta/M$. We are looking for the smallest integer $M$ which solves the inequality $\theta/M \le c(LM/d)$. This is
\be
    M = \left\lceil \sqrt{\frac{\theta d}{cL}} \right\rceil\,.
\ee
We tried various values for the constant of proportionality $c$.  Smaller $c$ lead to more accurate evolution but longer runtime, because the runtime is cubic in the number of segments.  We settled on $c = 4\times 10^{-3}$ because reducing $c$ below this level produces little change in the results below, particularly the spectrum $P_n$ and the strengths of cusps.

Note that since, in the piecewise-linear model, all segments are followed by ``kinks'' (in the sense of a discontinuous change in the tangent vector), we apply the splitting procedure to all segments without attempting to discriminate between ``physical'' and ``model'' kinks. However, this procedure is idempotent, and the majority of segments return $M=1$. Whenever we discuss the segmentation of a loop, we mean its segment count ``fresh from simulation'', i.e., before the splitting procedure has been applied. This splitting procedure is applied to each string only once, before starting to evolve it under backreaction.

\subsection{Loop evaporation fractions}\label{ssec:evaporation-fraction}

A final important question is: for how long should we evolve each loop? Certainly more time is better, but it comes at an increasing cost. In addition to studying backreaction in general, we want to be able to say something about a typical loop in the network. To set our benchmark, we therefore ask: what is the distribution of the fraction of length lost to gravitational backreaction in a network of loops?

Let $L_0$ be the loop's length at the time of creation $t_0$, and $L$ be its length at some later time $t$. The loop length changes in time as $L = L_0 - \Gamma G\mu t$ to first approximation, assuming $t\gg t_0$ and $\Gamma$, the total power emitted in gravitational waves by the loop, constant.\footnote{The arguments in this section may be repeated, keeping $t_0$, but doing so does not change the conclusion.} With $x=L/t$, we can rewrite this as $L_0 = L\left[1+(\Gamma G\mu/x)\right]$. Then, defining the \emph{evaporation fraction} of a loop as $\chi=1-L/L_0$, we find
\be
    \chi = \frac{\Gamma G\mu}{x+\Gamma G\mu}\,.
\ee
The distribution of loops as a function of their current and initial $x$ is~\cite{Blanco-Pillado:2013qja}
\be
    \textsf{n}(x,x_i) = \frac{(x_i+\Gamma G\mu)^{3/2}}{(x+\Gamma G\mu)^{5/2}}f(x_i)\,,
\ee
where $f(x_i)$ is the loop production function, with units of the number of loops per conformal time per conformal volume per $x$.  Integrating out $x_i$,
\be
    \textsf{n}(x) = \frac{\int_0^\infty(x_i+\Gamma G\mu)^{3/2}f(x_i)\,dx_i}{(x+\Gamma G\mu)^{5/2}}\,.
\ee
The relative distribution of loops in $x$ is then
\be
    \operatorname{Pr}(x) = \frac{\textsf{n}(x)}{\int^\infty_0 \textsf{n}(x)\,dx} = \frac 32\frac{(\Gamma G\mu)^{3/2}}{(x+\Gamma G\mu)^{5/2}}\,,
\ee
noting that all $x_i$ dependence has canceled out. We can convert this into a distribution on evaporation fraction using $\operatorname{Pr}(\chi)\,d\chi = \operatorname{Pr}(x)\,dx$ and $|dx/d\chi| = \Gamma G\mu/\chi^2$ to find
\be\label{eqn:Pr-ep}
    \operatorname{Pr}(\chi) = \frac32\chi^{1/2}\,,\qquad\operatorname{Pr}(\chi\geq \chi_*) = \int^1_{\chi_*} \operatorname{Pr}(\chi')\,d\chi' = 1-\chi_*^{3/2}\,.
\ee
Thus the distribution of loops is skewed toward those with larger evaporation fraction. In particular, 50\% of loops are at least $63.0$\% evaporated. In order to be able to say something about the majority of loops in a network, we set $\chi_*=0.7$ (70\% evaporated) as our benchmark, and we typically make comparisons in steps of 10\% evaporation.

Next, let's estimate the effort involved in evaporating to this level. Our step size is in units of $s \sim N_\text{osc}G\mu$, and so each iteration of backreaction represents some number of oscillations of a loop. But because a loop's period of oscillation is one-half of its length, the fractional change of a loop's length after the $i$th step of backreaction is
\be\label{eqn:dL-per-step}
    \frac{L_i-L_{i-1}}{L_{i-1}} = -\frac{\Gamma \Delta s_i}{2}\,,
\ee
with $\Delta s_i$ the step size set by the DOP853 routine at the $i$th step. This implies that for a fixed step size and $\Gamma$, the fractional change in length is fixed, and so an infinite number of steps is required to reach $L=0$.\footnote{The need for an infinite number of steps to reach $L=0$ does not result from anything specific to our procedures.  A loop loses a certain fraction of its length in each oscillation, and so undergoes an infinite number of oscillations before completely evaporating (in the limit of zero string thickness).} Loops with complicated structure can cause the code to lower $\Delta s_i$ until the structure is resolved, but conversely the step size can increase for smooth loops. Most of our loops evaporated to $\chi=0.7$ reach that point at around 4000 steps.

Since the computational effort for our code grows as $N_AN_B(N_A+N_B)$, it is not possible to evolve highly segmented loops to significant evaporation fractions on reasonable timescales, even with highly parallelized code on a high-performance cluster such as the one we used. Our first step to control the total segmentation of our loops is to look only at loops produced in the range $\tau=[250,400]$ in our network simulation: this prevents any one loop from being too large, and restricting ourselves to loops from a time range of the simulation (rather than a segmentation range of loops from the entire simulation) imposes a first cut based on a physical property rather than our ability to evolve a loop. From this starting pool of 198 loops, we proceeded as follows:
\begin{enumerate}
    \item All loops are evaporated to $\chi=0.1$ in order to understand how the spectra of loops change early in their lifetime, which is important for extrapolating to real loops.
    \item Prior work~\cite{Blanco-Pillado:2019nto} suggested that certain loop characteristics, such as $\Gamma$, stabilize when the loop reaches $\sim 50\%$ evaporation ($L=L_0/2$). We therefore evolve all loops for which neither the $A$ nor the $B$ representation had more than 500 segments to $\chi=0.5$. These 148 total loops are used for validation and extrapolation of other results.
    \item Finally, we evolve all loops for which neither the $A$ nor the $B$ representation had more than 300 segments to $\chi=0.7$. These 105 total loops, the ``70\% sub-population'', will provide much of the results to follow.
\end{enumerate}
A summary of these populations, such as their count and segmentations of member loops, can be found in Table~\ref{tbl:evaporations}.

\begin{table}
    \begin{tabular}{cccc}
        Maximum evaporation fraction & \# of loops & Range of \# segments & Mean \# segments\\\hline
        $0.7$ & 105 & $[187,582]$ & 369\\
        $0.5$ & 43 & $[499,922]$ & 658\\
        $0.1$ & 50 & $[852,3304]$ & 1594\\
    \end{tabular}
    \caption{The number of loops evaporated to various degrees, along with measures of their size (in terms of total segment count). All loops are taken from the same temporal range of the network simulation.}\label{tbl:evaporations}
\end{table}

In terms of total computation time, the $\chi=0.1$ sub-population took $\sim 3.3$ times more effort than the $\chi=0.7$ sub-population, and the $\chi=0.5$ sub-population took $\sim 1.3$ times more.

\section{Gravitational wave power}\label{sec:power}

Once we have found the shape of loops at various stages of
evaporation, we would like to compute the gravitational wave power
emitted at each stage.  In principle this could be done by looking at
the loss of length from gravitational back reaction according to
Eq.~(\ref{eqn:Xuv}), as described by Allen and Casper
\cite{Allen:1994iq}.  However, there are several disadvantages.
First, our computation approximates the backreaction on each segment
of $A$ or $B$ as the effect on its center.  In this technique we
would have to instead compute the total effect on each segment.
Second, this procedure calculates only the total radiation, not the
spectrum.  Finally, its runtime is cubic in the number of segments.
This is the same as the backreaction calculation, so it would not be
an obstacle for backreacted loops.  However, with a faster technique
we can calculate the gravitational wave power even for much larger
loops.  So we proceed as follows.

First, for harmonics up to $2^{14}$, we compute the gravitational wave power using the methods of \cite{Allen:2000ia,Blanco-Pillado:2017oxo}.  To find the radiation in a given direction $\ohat$, we first compute
\be\label{eqn:I}
\bI^{(n)}(\ohat) = \frac1L\int_0^L dv\,\bA'(v)
e^{(2\pi i n/L) (v - \ohat\cdot \bA(v))}
\ee
and similarly $\bJ^{(n)}(\ohat)$ in terms of $\bB(u)$.
Then using $\bI$ and $\bJ$, we can compute the power as
\be\label{eqn:dPdO}
\frac{dP^{(n)}}{d\Omega}
= n^2 (|I_\perp|^2|J_\perp|^2+ 4 \Im(I_xI_y^*)\Im(J_xJ_y^*))\,,
\ee
where $x$ and $y$ denote any two directions perpendicular to $\ohat$,
and $|I_\perp|^2 = |I_x|^2+|I_y|^2$ and similarly for $\bJ$.

We use a piecewise-linear form for $A$ and $B$, so we can write the tangent vector $A'$ as a sequence of constant pieces $A'_i$, with $A'_i$ in effect for $v=v_i$ to $v_{i+1}$.  The integral in Eq.~(\ref{eqn:I}) then becomes a sum \cite{Allen:2000ia},
\be\label{eqn:Isum}
\bI^{(n)}(\ohat)=\frac{1}{2\pi i n}\sum_{i=0}^{N_a-1}
\left[\frac{\bA'_{i-1}}{1-\ohat\cdot\bA'_{i-1}} -
\frac{\bA'_i}{1-\ohat\cdot\bA_i'}
\right]
e^{-2\pi i (n/L) (v_i-\ohat\cdot\bA_i)}
\ee
and similarly for $\bJ$.
Equation~(\ref{eqn:Isum}) is a nonuniform Fourier transform, which we
compute using the method of \cite{Potts}.

To get the total radiated power we need to integrate over solid angle.
We follow the same procedure as \cite{Blanco-Pillado:2017oxo} using
the idea of \cite{Atkinson:sphere}.  We divide the sphere into 81920
triangles with nearly identical area, evaluate the radiation spectrum
at the center of each, and multiply by the area to integrate the
radiation.

This method is $\mathcal{O}(N\log N)$ in the number of harmonics to be computed
and linear in the number of segments in the loop.  It gives good
performance up to around $10^5$ harmonics.  To extend the calculation
to much larger numbers, we would like a procedure which will give us a
binned spectrum without having to compute the power in every
individual harmonic.

To do that we first write Eq.~(\ref{eqn:Isum}) in the general form
\begin{subequations}\begin{align}
I_x^{(n)}&=\frac{1}{2\pi i n}\sum_{j=0}^{N_a-1} c^x_j e^{-2\pi i n p_j}\\
J_x^{(n)}&=\frac{1}{2\pi i n}\sum_j d^x_j e^{-2\pi i n q_j}
\end{align}\end{subequations}
and similarly for $I_y$ and $J_y$.
We now write
\be\label{eqn:Iperp4}
|I_\perp^{(n)}|^2=\frac{1}{4\pi^2 n^2}
\sum_{jj'} e^{2\pi i n (p_j-p_{j'})} (c^x_j c^x_{j'} + c^y_j c^y_{j'})
\ee
and similarly for $J$, so
\be
\frac{dP}{d\Omega} \supseteq 8 \pi G\mu^2 \sum_{n=1}^\infty
\frac{1}{16\pi^4 n^2}
\sum_{jj'}\sum_{kk'}
e^{2\pi i n (p_j-p_{j'}-q_k-q_{k'})} 
(c^x_j c^x_{j'} + c^y_j c^y_{j'})(d^x_k d^x_{k'} + d^y_k d^y_{k'})
\ee
This is only the $|I_\perp|^2|J_\perp|^2$ term.  The
$\Im(I_xI_y^*)\Im(J_xJ_y^*)$ term could be computed similarly.

The advantage of this is that we can now do the sum over $n$
\cite{Allen:2000ia}.  The summand is complex conjugated by the
exchange $j\leftrightarrow j',k\leftrightarrow k'$, so the exponential
can be replaced by a cosine, and then we can use
\be
\sum_{n=1}^\infty \frac{\cos n\theta}{n^2}
 = \frac{\pi^2}{6}-\frac{\pi\theta}{2} +\frac{\theta^2}{4}
\ee
for $\theta\in[0, 2\pi]$ to get the total power.  To get a binned
spectrum, we can use
\be\label{eqn:Lerchsum}
\sum_{n=n_1}^\infty \frac{\cos n\theta}{n^2}
= \Re \left[e^{2\pi i n_1 x} \zeta(\theta, n_1, 2)\right]
\ee
where $\zeta$ is the Lerch $\zeta$ function,
\be
\zeta(x, a, s) = \sum_{k=0}^\infty \frac{e^{2\pi i k x}}{(k+a)^s}\,.
\ee
Thus 
\be\label{eqn:Lerchsum2}
\sum_{n=n_1}^{n_2} \frac{\cos n\theta}{n^2}
= \Re \left[e^{2\pi i n_1 x} \zeta(\theta, n_1, 2)
- e^{2\pi i n_2 x} \zeta(\theta, n_2, 2)\right]\,.
\ee

The problem with this technique is that it takes time of order
$N_a^2N_b^2$, and thus is intractable.

Instead of the above exact calculation that is quartic in the number of
segments, we calculate the harmonics above $2^{14}$ with an
approximate quadratic calculation.  First, consider the two terms in Eq.~(\ref{eqn:dPdO}).  The first is positive definite. If we average over many loops, it will contribute to the average spectrum.  But the second term can have either sign, depending on the detailed phases of the components of $I$ and $J$.  We expect that the contribution from this term to the overall spectrum will be very small because of cancellation.  In numerical experiments, the contribution of the second term to the power in harmonics $2^{14}\ldots 2^{15}$ is only around $10^{-4}$ times than that of the first term.

Considering only the first term in Eq.~(\ref{eqn:dPdO}), we define
\be\label{eqn:Cn}
C_n = n|I_\perp^{(n)}|^2 = 
\frac{1}{4\pi^2 n}
\sum_{jj'} \cos(2\pi n (p_j-p_{j'})) (c^x_j c^x_{j'} + c^y_j c^y_{j'})
\ee
and similarly $D_n = n|J_\perp^{(n)}|^2$.  Then
\be
\frac{dP}{d\Omega} \approx 8 \pi G\mu^2 \sum_{n=1}^\infty
C_n D_n
\ee

Suppose $C_j$ and $D_j$ had some constant values $C$ and $D$ everywhere in
a bin.  Then the total power in a bin would be
\be\label{eqn:IJconst}
8 \pi G\mu^2
n_{\text{bin}} C D
\ee
where $n_{\text{bin}}$ is the number of frequencies in the bin.  In fact
$|I_\perp|^2$ and $|J_\perp|^2$ are not constant at all, but have
large fluctuations.  However, these fluctuations depend on the exact details of the functions $A$ and $B$ and are generally not
correlated between $I$ and $J$.  So let $C$ be the average of the $C_j$ in a bin and $c_j$ be the deviation from the average, so $C_j = C + c_j$, and
$D= D + d_j$ similarly.  The sums over a bin of $c_j$ and $d_j$ vanish by definition, and the sum of $c_jd_j$ vanishes in an ensemble average, since the fluctuations are uncorrelated.  Since we are summing over a large number of elements in each bin, this term makes very little contribution in almost all realizations.  Thus
\be\label{eqn:CD2}
\sum_j C_n D_n \approx n_{\text{bin}} C D\,, 
\ee
which we can compute using Eqs.~(\ref{eqn:Cn}) and (\ref{eqn:Lerchsum2}).

In addition to the fluctuations, there may a correlated, secular change in $C$ and $D$ as the spectrum changes across the width of the bin, which will cause a deviation from Eq.~(\ref{eqn:CD2}).  Suppose $C_j, D_j \sim j^\alpha$.  We have many frequencies in one bin, so we replace these with continuous functions $C(f) \sim f^\alpha$, and similarly for $D$.   
In a bin from frequency $f_1$ to $f_2$ we can write $C(f) = (f/f_1)^\alpha C(f_1)$, so the average is
\be\label{eqn:Cbar}
\bar C = \frac{1}{f_2-f_1}\int_{f_1}^{f_2} df C(f) = \frac{s^{\alpha+1} -1}{1+\alpha}
\frac{C(f_1)}{s-1}
\ee
with $s=f_2/f_1$.  The average of $D$ is similar, but when we average $CD$, we get
\be\label{eqn:CDbar}
\overline{CD} = \frac{s^{2\alpha+1} -1}{s-1}
\frac{C(f_1)D(f_1)}{1+2\alpha}
\ee
so
\be\label{eqn:CDratio}
\frac{\overline{CD}}{\bar C \bar D}= \frac{s^{2\alpha+1} -1}{1+2\alpha}
\left(\frac{s^{\alpha+1}-1}{1+\alpha}\right)^{-2}(s-1)
\ee
If $\alpha = -1/2$, the first fraction in Eq.~(\ref{eqn:CDratio}) should be replaced with $\ln s$, and if $\alpha = -1$ the second fraction should be replaced with $\ln s$.  The result is still continuous.

In the results below, we find that the slope of the GWB is generally between $-2$ and 2, so $-1 \le\alpha\le1$.  If we use Eq.~(\ref{eqn:CD2}) with $s = 2$, Eq.~(\ref{eqn:CDratio}) tells us that we may make an error of as much as about 4\%.  To avoid this, we do the above calculation with each bin split into 4 pieces, and combine them afterwards.  Then $s = 2^{1/4}$ and the maximum error is reduced to about 0.25\%.

In fact it is not necessary to do the full quadratic sum in
Eq.~(\ref{eqn:Cn}).  We only use this technique for large $n$, in
which case rapid oscillation of the cosine leads to a tiny
contribution unless $p_j-p'_j$ is small.  Thus for each $j$, we
consider only nearby $j'$ where
\be\label{eqn:Cthreshold}
|p_j-p_{j'}| < \frac{N_A}{2\pi n}\,.
\ee

The threshold is computed as follows.  We are trying to compute the
sum of $C_n$ in a bin from $n_1$ to $n_2$.  Let's suppose that the
second factor in Eq.~(\ref{eqn:Cn}) is about 1.  Then each $j,j'$ pair
gives rise to a term of order
\be\label{eqn:Csum1}
\sum_{n_1}^{n_2} \frac{\cos(x n)}{n}\,,
\ee
where $x = 2\pi(p_j-p_{j'})$.  We're considering the case where $x \ll
1$, so the sum can be approximated by an integral,
\be
\int_{n_1}^{n_2} dn \frac{\cos(x n)}{n} = \Ci(n_2 x) - \Ci(n_1 x).
\ee
When $n_1 x \ll 1$, this is about $\ln n_2/n_1$, which is of order 1.
When $n_1 x \gg 1$, we can use the asymptotic approximation $\Ci(z)
\sim \sin z/z$, so $|\Ci(z)| \alt 1/z$, and Eq.~(\ref{eqn:Csum1}) is
of magnitude at most $1/(xn)$.

There are $N_a$ values of $j$, and thus $N_a^2$ pairs $j,j'$.  But
when we add these up, the signs are essentially random, so the total
contribution is of order $N_a/(x n)$.  Setting this less than 1
yields Eq.~(\ref{eqn:Cthreshold}).  The situation with $D$ in terms of
$N_b$ is analogous.  We can check whether the threshold is sufficient
by increasing it to include more $j'$ values; this makes no
significant difference in the result.

For large $n_1$, only a few $j'$ values need to be considered.  So
while this method is formally quadratic in $N_a$ or $N_b$, in practice
it is essentially linear, and thus enables a fast calculation of the
binned gravitational wave spectrum at arbitrarily high harmonics.  We
can check whether this method is accurate by comparing its result for
intermediate $n$ to the result from the FFT-based calculation.
Indeed we find it does a good job of computing the power with much
less runtime.

\section{Results}\label{sec:results}

\begin{figure}
    \centering
    \includegraphics[scale=0.50]{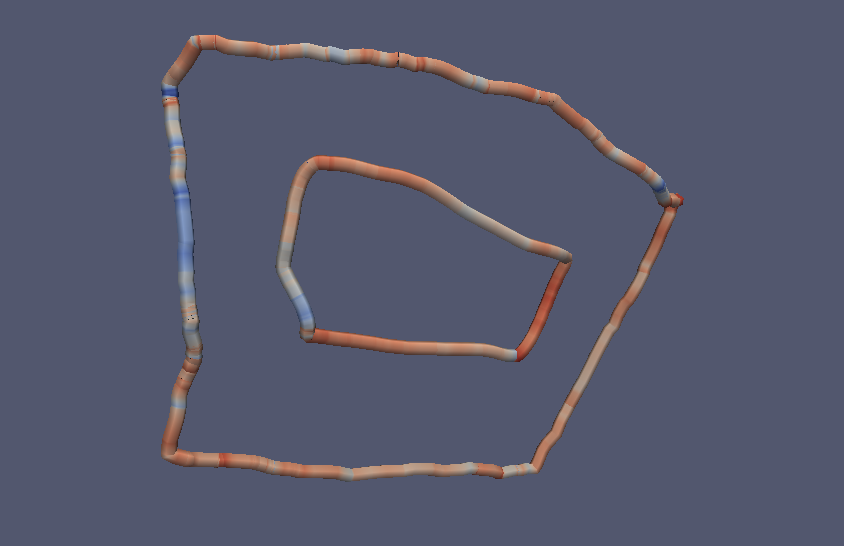}
    \caption{An example of how backreaction changes a loop for a single member of our corpus shown at 0\% evaporation and 50\% evaporation. This comparison is done at corresponding points in the oscillation period; bluer segments are those more slowly moving at that instant, and redder segments more quickly. The unevaporated loop has a more jagged structure, both in the sense of directional variation along the loop and in the sense of changing speed from segment to segment. The evaporated loop retains the general rectangular shape, with four large kinks, but both the directional and speed variations have been smoothed out by backreaction. This loop stays in a non-self-intersecting trajectory throughout the entire process of backreaction, so the number of kinks stays constant during its evaporation.}\label{fig:loop-comparison}
\end{figure}

Gravitational backreaction generically changes the shape of loops; the only known exceptions are loops with maximal symmetry (e.g., the ACO loop~\cite{PhysRevD.50.3703,0264-9381-22-13-002}). Almost all potentially observable signals from loops depend on their shape. One of the principal effects of backreaction is to smooth out the small-scale structure on loops; qualitatively, they become less ``jagged'', as shown for a representative member of our corpus (Loop \#152) in Fig.~\ref{fig:loop-comparison}.
This removal of small-scale structure can also be seen quantitatively, which we discuss in more detail in Secs.~\ref{ssec:Pn} and \ref{ssec:kinks} below. Other effects of backreaction include causing minor self-intersections in previously non-self-intersecting loops (Sec.~\ref{ssec:intersections}) and producing very weak cusps by the filling-in of kinks (Sec.~\ref{ssec:cusps}).

For much of the following analyses, we restrict ourselves to studying the 105 loops which reached $\chi=0.7$, the 70\%-evaporated sub-population. This is because we wish to see how various measures of the loops change with $\chi$, and using different sets of loops (with different segmentations!) at different $\chi$ could bias the results. When we only analyze results out to smaller $\chi$, we include all loops which reached at least that threshold, and draw attention to this fact in the text.

\subsection{Self-intersections}\label{ssec:intersections}

Prior toy models of backreaction observed self-intersections of strings due to backreaction, although the length lost was minor~\cite{Blanco_Pillado_2015}. Our numerical backreaction also sees small length losses in regions where the $A'$ and $B'$ cross, but more rarely larger length losses as well. In a few cases, backreaction can lead to the ``fragmentation'' of a loop, where a single loop breaks into multiple smaller loops, the largest of which is only a few tenths of the original loop's size. However, the general trend is that self-intersections are quite generic, but typically minor. 

In order to find intersections, we simulate one oscillation of each loop after completing each step of $10^{-5}$ in $s$, recording relevant statistics such as the number of intercommutations and the length lost to self-intersections (if any are found). In the event of self-intersections, at least two loops exist at the end of the simulation. In effectively all cases, one loop is larger by a significant factor; we keep this longest loop as the ``true'' loop which we're evolving and discard the small ``looplet(s)'' produced by the self-intersection.

We divide the 105 loops in the 70\%-evaporated sub-population into two nonoverlapping categories: those with \emph{major} length losses due to self-intersections, and those with \emph{non-major} length losses due to self-intersections. Our definition of \emph{major} is that the loop loses more than 5\% of its initial length to self-intersections over the course of its evolution. This is chosen because we examine our loops at fixed evaporation fractions in steps of $0.1$ (or 10\%), and so a ``non-major'' loop at any $\chi$ is guaranteed to have at least $\chi-0.05$ of its length loss be due to backreaction. The loops without any intersections clearly have exactly $\chi$ of their length lost due to backreaction,\footnote{In practice, we use the first multiple of $10^{-5}$ in $s$ for which the loop's evaporation fraction exceeds a particular $\chi$ when calculating loop quantities reported at $\chi$. Since our steps in $s$ are $\ll1$, this difference is negligible, per Eq.~(\ref{eqn:dL-per-step}).} and so if a loop loses more than $0.05L_0$ of its length to self-intersections, we might expect it at some $\chi$ to be more similar to the no-self-intersections loops at the prior slice, $\chi-0.1$.

\begin{figure}

    \centering

    \subfloat[\label{fig:isexn_total-length-lost}]{%
        \includegraphics[width=0.4\textwidth]{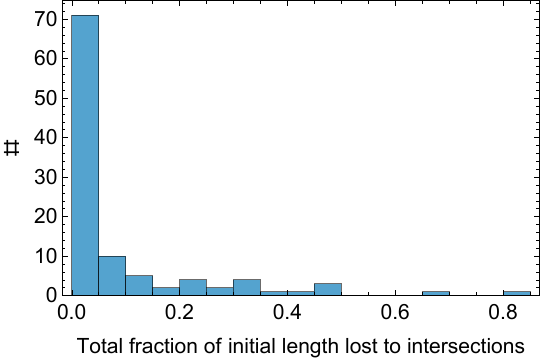}
    }
    \hspace{1em}
    \subfloat[\label{fig:isexn_total-isexns}]{%
        \includegraphics[width=0.4\textwidth]{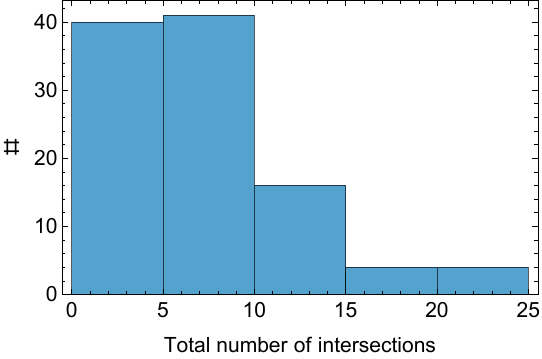}
    }

    \subfloat[\label{fig:isexn_frac-length-lost}]{%
        \includegraphics[width=0.4\textwidth]{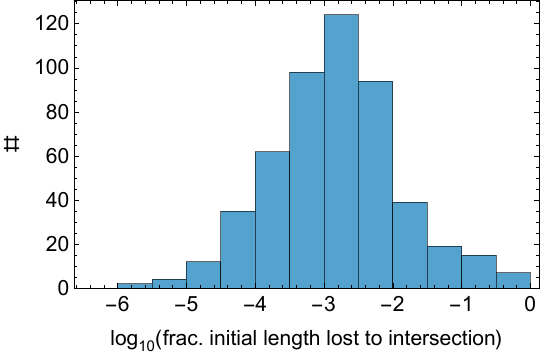}
    }
    \hspace{1em}
    \subfloat[\label{fig:isexn_recoil-v}]{%
        \includegraphics[width=0.4\textwidth]{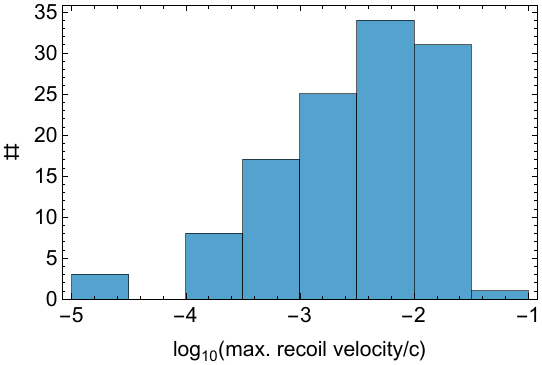}
    }

    \subfloat[\label{fig:isexn_ep-dist}]{%
        \includegraphics[width=0.4\textwidth]{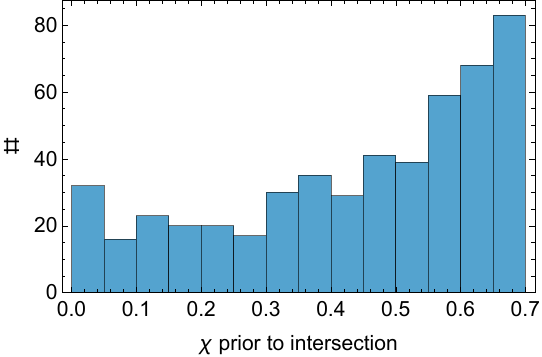}
    }
    \caption{Distributions of various measures of intersections due to gravitational backreaction for the 70\%-evaporated sub-population. While intersections are generic, they are generally small in terms of the length lost as well as their effect on the motion of the loop. While intersections become more common with increased evaporation fraction, there is no correlation between $\chi$ and the amount of length lost to intersection.}\label{fig:isexn}
\end{figure}

With this definition, we find 71 (68\%) of loops do not experience major length losses and 34 (32\%) of loops do. Of the 71, 9 experience no self-intersections at all. The 71 loops form an important sub-population we term the \emph{no-majors} subpopulation; we use it repeatedly for studying how various properties of the loops change with evaporation fraction. It should be emphasized that even for loops not in the no-majors subpopulation, the total length lost to intersections is not typically extreme; the 80th percentile of total length lost to intersections is $0.11L_0$, and more than half of the initial length is only lost for two loops ($0.688L_0$ and $0.817L_0$). Thus, fragmentation of loops due to gravitational backreaction is rare. The distribution of total length lost to intersections is shown in Fig.~\ref{fig:isexn_total-length-lost}.

Checking the 70\%-evaporated sub-population after each step of  $10^{-5}$ in $s$, we find 512 such simulations which contain at least one intersection, and 676 intersections in total; most simulations with intersections contain only one (445, or 87\%), but the distribution has a very long tail, with one simulation containing 13 intersections. Most loops (85\%) experience more than one intersection over the course of their evolution, with the distribution of total intersections for the loops shown in Fig.~\ref{fig:isexn_total-isexns}.

The typical intersection leads to very little length loss, as can be seen in Fig.~\ref{fig:isexn_frac-length-lost}; the geometric mean of the loss to intersection is $10^{-2.9}L_0$, and $94\%$ of intersections result in a loss of less than $0.05L_0$. This works out to 32 intersections with loss of greater than $0.05L_0$, indicating that most of the loops which experience major length losses (of which there are 34) do so in a single intersection event.

Any intersection leads to a recoil velocity for all loops produced, as a result of momentum conservation; the distribution of the largest of these recoil velocities for all loops is shown in Fig.~\ref{fig:isexn_recoil-v}. The largest velocity found, in units of $c$, is $0.296$, and $64\%$ of loops experience a maximum recoil velocity of greater than $10^{-3}$. As a result, self-intersections are a potential mechanism for unbinding loops from galaxies, where they might otherwise cluster by gravitational interactions~\cite{Chernoff:2017fll,Jain:2020dct}. This has implications for studies of the rocket effect~\cite{Durrer:1989zi,Casper:1995ub}, gravitational lensing by strings, and the detectability of bursts from strings~\cite{Yu:2014gea,Chernoff:2017fll,Auclair:2023mhe}.

Finally, we can ask when in a loop's lifetime an intersection is likely to occur. All of our loops are initially non-self-intersecting, and so we might expect for the likelihood of intersection to increase with $\chi$, as backreaction has more time to move the loops onto self-intersecting trajectories. This is shown in Fig.~\ref{fig:isexn_ep-dist}. After an initial spike of intersections, likely due to loops created with nearly-self-intersecting trajectories, we find intersections happen at a fairly low rate until around $\chi=0.3$, at which point they rise in frequency. However, our data indicate that there is no correlation between the evaporation fraction at which an intersection occurs and how much length is lost to that intersection.

\subsection{Changes to the loop power spectrum}\label{ssec:Pn}

For gravitational wave detections, understanding the loop power spectrum, $P_n$ (with $n$ the mode number), is of primary importance. For example, the gravitational wave background (GWB) is found from~\cite{Blanco-Pillado:2017oxo}
\be\label{eqn:GWB}
    \Omega_\text{gw} \propto \sum_n C_nP_n\,,
\ee
where $C_n$ is a coefficient depending on the distribution of loop sizes in time and space as well as cosmological history. However, since gravitational backreaction affects the shape of the loop, the GWB also needs to know the distribution of $\chi$ values of loops in the network at any given time as well as the dependence of $P_n$ on $\chi$. We focus here on the latter, using the no-majors population (Sec.~\ref{ssec:intersections}) for all results.

\begin{figure}
    \centering
    \includegraphics[scale=1.0]{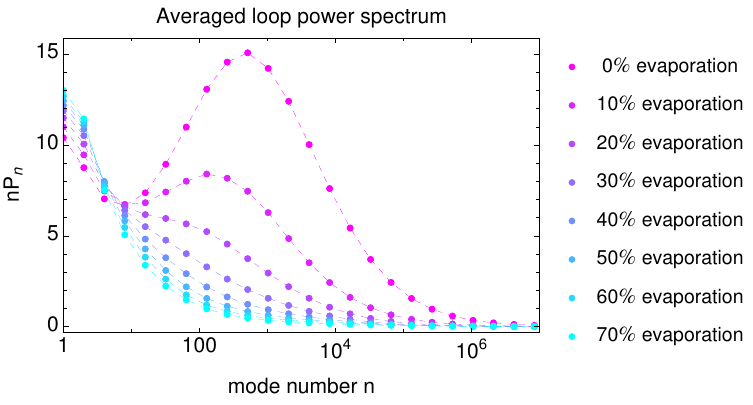}
    \caption{The change to the average power spectrum, $P_n$, with evaporation fraction for the 70\%-evaporated subpopulation with all loops given equal weight. The shape of the spectrum at low mode numbers, representing the large-scale structure of the loop, is not greatly changed. The significant change at moderate mode numbers is due to the smoothing of small-scale structure. The reduced amplitude of the bump indicates that the typical size of a change in the string direction is decreasing, and the reduced value of $n$ for this peak indicates that the length scale on which we expect a change in direction is increasing.}\label{fig:Pn-by-evaporation}
\end{figure}

The change to the power spectrum with evaporation fraction can be seen in Fig.~\ref{fig:Pn-by-evaporation}.
The most notable feature here is the large bump at moderate values of $n$.  (However, note that the vertical axis here is $nP_n$, as appropriate for $\log n$ on the horizontal axis.  Thus it is not that the individual $P_n$ are large in this range but that the total power over each logarithmic interval of $n$ is large.). This bump is reduced by backreaction and effectively vanishes by $\chi=0.3$.

This vanishing of the bump is indicative of the smoothing of small-scale structure. For a string of length $L$, the value of $P_n$ for any $n$ captures the typical contribution of a segment of string of length $L/n$ to the power spectrum. Thus, a peak at some $n_\text{max}$ indicates a strong contribution at a scale of $L/n_\text{max}$ on the loop, meaning that we should expect significant structure (changes in the direction of the string) at those scales on the loop. The decrease in the height of the peak indicates that the variation in this structure (or the typical magnitude of a change in direction) is decreasing; loosely, the string is becoming smoother. The decreasing value of $n_\text{max}$ indicates that the scale on which we expect variation is growing.

By the time we reach about $\chi=0.3$, the bump has vanished. This is not to say that the loop is completely smooth; it's still possible to have a few large, rapid changes in the direction of the loop without leading to a large bump in the power spectrum at the associated mode number.\footnote{Alternately, one may have such changes at all length scales, as in the ``pure kink'' power spectrum commonly used in predicting cosmic string GWB, which follows a power-law in $n$.} Our loops typically have a few of these large, rapid changes, even at $\chi=0.7$. We discuss the implication of this for cusps further in a following section. As the $nP_n$ represented here are for the least-segmented subpopulation we studied, more segmented loops (formed, on average, at later times) will have larger bumps at low evaporation fraction, although these are still quickly removed by backreaction (see Sec.~\ref{ssec:kinks} for further discussion). However, the impact on the GWB would be reduced, as loops at greater $\chi$ contribute more than those at lesser $\chi$; see Ref.~\cite{sgwb-paper} for further details.

As a final note regarding $P_n$, its values at the smallest $n$ do not significantly change shape with $\chi$, although the amplitude increases by around 20\%. These smallest $n$ represent the largest-scale structure, or what one might call the overall ``shape'' of the loop. The similar shape can be understood as a consequence of the scale at which backreaction is effective growing as $\Gamma G\mu t$. With $\Gamma G\mu\ll 1$, we would need to wait until extremely late times until the entire loop's shape is significantly modified by backreaction, on the order of $t\sim L_0/\Gamma G\mu$. However, this is comparable to the loop lifetime, and so in effect the loop's initial shape persists for most of its life.

The data shown in Fig.~\ref{fig:Pn-by-evaporation} is available on \href{https://doi.org/10.5281/zenodo.14037540}{Zenodo}~\cite{wachter_2024_14037540}.

Another measure of interest related to the power spectrum is the measure of the total power radiated in gravitational waves,
\be\label{eqn:Gamma}
    \Gamma = \sum_n P_n\,.
\ee
This tells us roughly how much energy loss (length loss) to expect per unit time for the loop; accounting for the coupling to gravity gives $dL/dt = -\Gamma G\mu$, from which we obtain the common approximations $L = L_0 - \Gamma G\mu t$ (valid for $\Gamma$ constant in time) and $\Delta L = \Gamma G\mu L/2$ (for a single oscillation of period $T=L/2$ at constant $\Gamma$). 

Studies of loop populations produced in simulations have found an average value of $\Gamma\sim 50$~\cite{Blanco-Pillado:2017oxo}, and it is this value, taken as constant, which is typically used in making predictions about gravitational waves from loops. However, previous results~\cite{Blanco-Pillado:2019nto} found that loops with $\Gamma$ greater than the canonical value have their $\Gamma$ reduced by backreaction; because these loops form with lots of small-scale structure, and thus a larger $\Gamma$ than the canonical one, we should instead expect the typical loop $\Gamma$ to change over time.

\begin{figure}
    \centering
    \includegraphics[scale=1.0]{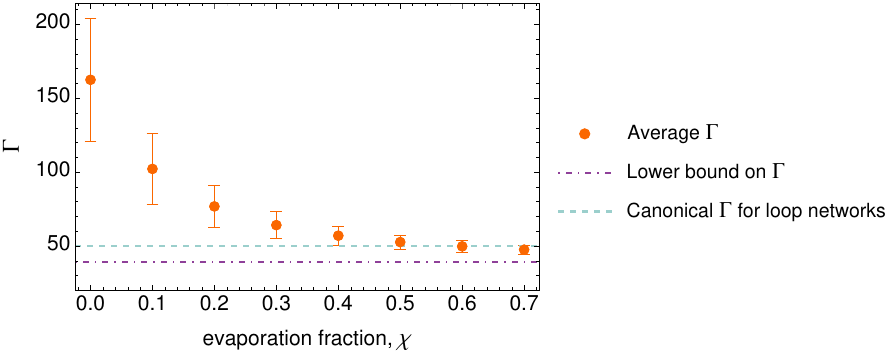}
    \caption{The change to the rate of gravitational energy emission, $\Gamma$, with evaporation fraction for the 70\%-evaporated subpopulation. While loops initially have large $\Gamma$ due to small-scale structure, this is quickly removed by gravitational backreaction. The error bars are at the $1\sigma$ level; the distribution of $\Gamma$ at each $\chi$ skews towards smaller values. In all cases the minimum $\Gamma$ stay above the conjectured lower bound~\cite{PhysRevD.50.3703,0264-9381-22-13-002}.}\label{fig:Gamma-by-evaporation}
\end{figure}

This expectation is borne out in our results, as shown in Fig.~\ref{fig:Gamma-by-evaporation}.
The initial large $\Gamma$ is due to the small-scale structure on freshly-formed loops, and the reduction in $\Gamma$ with increasing $\chi$ comes mainly from this small-scale structure being smoothed out; some effect is due to the amplitude decrease at the very highest mode numbers, but this comparably moderate change cannot account for the reduction in $\Gamma$ by a factor $\sim 3.4$.

In addition to the average decreasing, the range of $\Gamma$ values becomes tighter with increasing $\chi$. No loop's $\Gamma$ decreases below the conjectured lower bound, $\Gamma_\text{min}\approx 39$~\cite{PhysRevD.50.3703,0264-9381-22-13-002}; the smallest value we observe, for $\chi=0.7$, is $43.7$. Very rarely, $\Gamma$ may increase with $\chi$; this was seen in $\lesssim1\%$ of the data used in Fig.~\ref{fig:Gamma-by-evaporation}, and the largest such positive change was by a factor $\sim 1.04$. The average $\Gamma$s for the most-evaporated loops are close to but in the case of $\chi=0.6$ and 0.7 slightly less than the canonical $\Gamma\approx 50$ value for a loop network. The data shown here appears to be approaching an asymptotic mean $\Gamma$ in the mid-40s.

We will discuss the impact of backreaction's changes to the spectrum on the gravitational wave background produced by a loop network in a companion paper~\cite{sgwb-paper}. An important consideration there is that the evaporation fraction $\chi$ is in proportion to the age of the loop if and only if $\Gamma$ is constant. Because we know that $\Gamma$ changes, rather significantly, loops will reach evaporation fractions earlier than the corresponding fraction of their lifetime (e.g., a loop with large initial $\Gamma$ might reach 10\% evaporation at only 1\% of its lifetime). We will discuss these effects in more detail in~\cite{sgwb-paper}.

\subsection{The formation of cusps}\label{ssec:cusps}

Our prior computational~\cite{Blanco-Pillado:2019nto} and analytic~\cite{Blanco-Pillado:2018ael} results suggest that loops which initially have kinks but not cusps will form cusps due to the ``filling-in'' of kinks by gravitational backreaction. This process replaces a jump from one point to another on the Kibble-Turok sphere with a smooth path that can cross another such path, causing a cusp. Technically, the original kinks are immediately replaced by smoother regions, but only on a very short length scale; far away, the structure still appears kink-like. The timescale at which a kink in either worldsheet function appears smoothed to a (null) distance $w$ is $t \approx L(G\mu H)^{-1} (w/L)^{1/3}$ (where the constant of proportionality $H$ depends on the average curvature of the string and is typically about 20)~\cite{Blanco-Pillado:2019nto}.

An immediate consequence of this is that loops which initially have kinks but not cusps will never form particularly strong cusps: the length of string involved in the cusp at the end of the loop's lifetime, $t_\text{life}\approx L/\Gamma G\mu$, will be $w \approx (20/\Gamma)^3L$; for the conjectured lower bound of $\Gamma_\text{min}\approx 39$, this works out to $w\approx 0.13L$ as an upper bound. Since we know loops form with $\Gamma>\Gamma_\text{min}$, most cusps will involve much less of the string. Compared to the typical assumption in calculating the strength of cusps on cosmic string loops---that the entire loop length $L$ is involved in the cusp---this would lead us to predict cusps formed due to backreaction are very weak indeed. However, the cubing here means that small variations in either the constant of proportionality or $\Gamma$ can meaningfully change the prediction. Let's see how this works out for our loops.

For the simple loop models we previously studied computationally, the question of when and where cusps formed was fairly straightforward and could be done by visual inspection. For the loops we currently study, it is less straightforward, and the large size of our corpus (loop count $\times$ steps of backreaction) makes visual inspection infeasible. Instead, we consider any crossing of the $A'$ and $B'$ on the unit sphere to be a cusp and calculate the associated cusp strength. While this leads to a number of spurious (and very weak) cusps, it also lets us set a stricter upper bound on the strength of cusps on our loops.

To get an idea of how we make this comparison, let us consider the power spectrum of a cusp, which falls as $n^{-4/3}$. From a loop's worldsheet functions, we can calculate the total coefficient of $n^{-4/3}$ from all cusps, which we'll call $Q$, following the procedure laid out in Appendix~\ref{app:cusp-calculation}. This procedure assumes that the energy (length, $\sigma$) in the segments adjoining a `cusp' is evenly distributed across the crossing of $A'$ and $B'$. This provides an overestimate on the strength of the cusp in the following sense (see~\cite{Blanco_Pillado_1999} for further details). The strength of a cusp is inversely proportional to $A''$ and $B''$, the rate of change of the tangent vectors. A uniform distribution of energy leads to a uniform $A''$ and $B''$; or, the interpolated $A'$ and $B'$ are changing at the same rate everywhere nearby the cusp, and so regardless of where the $A'$ and $B'$ cross, the cusp strength is the same. If the distribution of energy is uneven, most of the motion on the unit sphere takes place over a small range of the null coordinate where $A''$ and $B''$ are large. (For example, the tangent vector may linger near some point on the Kibble-Turok sphere before quickly jumping somewhere else on the sphere.) The crossings that lead to cusps are very likely to take place in these regions of large $A''$ and $B''$, leading to weak cusps.

In order for the actual cusp strength to be greater than the uniformly-distributed estimate, the $A'$ and $B'$ would have to cross when both $A''$ and $B''$ are small (as the product $|A''||B''|$ is our measure of interest), which is unlikely because these regions occupy little of the sphere. Thus, since a real string would have non-uniformly-distributed energy, we usually assign a cusp a greater strength than it would actually possess. Our cusp calculations which follow should therefore be treated as an (approximate) upper bound, as they overestimate the contribution of cusps to the gravitational-wave spectrum on two counts.

Having calculated $Q$, we can estimate the upper bound on the total fraction of the loop's $\Gamma$ which can be said to be due to cusp-like behavior. Visually, we can plot the actual power spectrum of any loop against the line $Qn^{-4/3}$ to see where, if anywhere, the upper bound on cusp-like behavior is a good match for the actual spectrum (certainly at low mode numbers, where the bulk structure of the loop is known to dominate in toy models of backreaction~\cite{Blanco-Pillado:2017oxo}, we expect some divergence).

\begin{figure}
    \centering
    \includegraphics[scale=1.0]{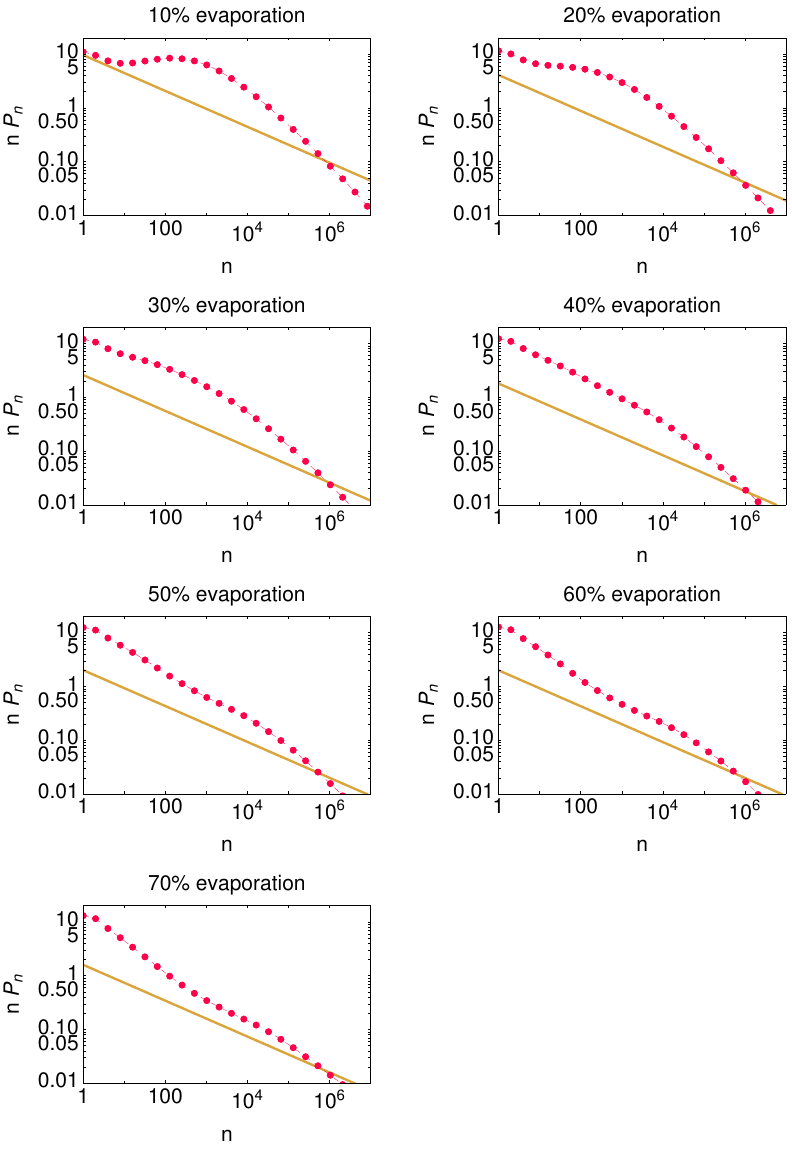}
    \caption{The change, with evaporation fraction, in the averaged upper-bound cusp-like spectrum of loops (gold lines) compared with the averaged actual spectrum of loops (red points). At all evaporation fractions, the total power in cusps is below the power in loops by a significant amount; at no points do we see the cusp line running along the actual spectrum at high frequencies, as we would expect for a loop whose GW emission is dominated by cusps.}\label{fig:cusp-comparison-all}
\end{figure}

This visual comparison is done for the averaged spectra of all loops in the 70\%-evaporated sub-population in Fig.~\ref{fig:cusp-comparison-all}.\footnote{The averaged spectra shown here are the same data as in Fig.~\ref{fig:Pn-by-evaporation}; we have plotted here on log-log axes to emphasize the power-law nature of the upper-bound cusp-like spectra.}
For loops whose radiation is dominated by cusps, we would expect the upper-bound cusp-like spectra to closely agree with, or ideally exceed, the actual spectra at mode numbers in the middle of the range we study. This is not the case for our loops at any evaporation fraction; the maximum contribution of cusps is sub-dominant up until high mode numbers, $n\gtrsim 10^5$, where the piecewise-linear nature of our loops means that their spectra begin to fall as\footnote{In a real loop, backreaction would smooth the kinks, as we discuss below.  Thus the $n^{-2}$ spectrum would be replaced with one that declines exponentially.  Here, there will always be kinks, because we can represent only a finite number of segments.  Thus the $n^{-2}$ spectrum will always be present, though at a reduced amplitude because of our introduction of additional points which allow the kink angle to decrease.} $n^{-2}$. The upper-bound averaged cusp coefficient decreases with $\chi$, both in absolute terms and when calculated as a share of the total power emitted into gravitational waves by the loops. The lines do run parallel at some evaporation fractions (e.g., $\chi=0.5,0.6,0.7$), but the cusp lines are still at best a factor $\sim 2$ below the actual spectra; as this is an upper-bound estimation, the actual cusp lines would be lower still.

At very high mode numbers, $n\gtrsim 10^6$, the upper-bound cusp lines would dominate on a real loop, which is not piecewise-linear. However, by the time we reach this point, the power spectrum is so low that almost none of the loop's total power can be said to be due to cusps. Assume that the real spectrum can be taken to be $\operatorname{max}(cusp,actual)$ and find that curve's $\Gamma$. The upper bound of the contribution due to cusps is this value minus the $\Gamma$ found just by integrating the actual power spectrum, and is summarized in Table~\ref{tbl:cusp-contribution}.

\begin{table}
    \centering
    \begin{tabular}{|c|c|c|c|c|c|c|c|}
        \hline
        $\chi$ & $0.1$ & $0.2$ & $0.3$ & $0.4$ & $0.5$ & $0.6$ & $0.7$ \\
        \hline
        $\Gamma$ w/ $P_n$ only & $102.1$ & $76.77$ & $64.11$ & $56.97$ & $52.62$ & $49.75$ & $47.51$ \\
        \hline
        $\Gamma$ w/ $P_n+$cusp & $102.4$ & $76.86$ & $64.17$ & $57.01$ & $52.68$ & $49.80$ & $47.55$ \\
        \hline
        \% diff. & $0.236$ & $0.126$ & $0.0914$ & $0.0684$ & $0.106$ & $0.101$ & $0.0773$\\
        \hline
    \end{tabular}
    \caption{The difference in total $\Gamma$, across evaporation fraction $\chi$, if we either: sum only the averaged $P_n$ calculated from the loops directly; or: sum the larger of either the averaged $P_n$ or the predicted cusp line. Because the cusp line dominates only at large $n$ and thus low $P_n$, the difference is minor, and so cusps don't significantly contribute to loops' gravitational-wave emission.}\label{tbl:cusp-contribution}
\end{table}

Neither Fig.~\ref{fig:cusp-comparison-all} nor Table~\ref{tbl:cusp-contribution} contain information about the 0\% evaporation loops. This is because we assume the shape of a loop taken from simulation to be its true shape, with all steps in $A'$ and $B'$ representing kinks, but at any later times allow for backreaction to have ``filled in'' the kinks, producing cusps.

\begin{figure}

    \centering

    \subfloat[][Kibble-Turok loop, no backreaction]{%
        \includegraphics[width=0.4\textwidth]{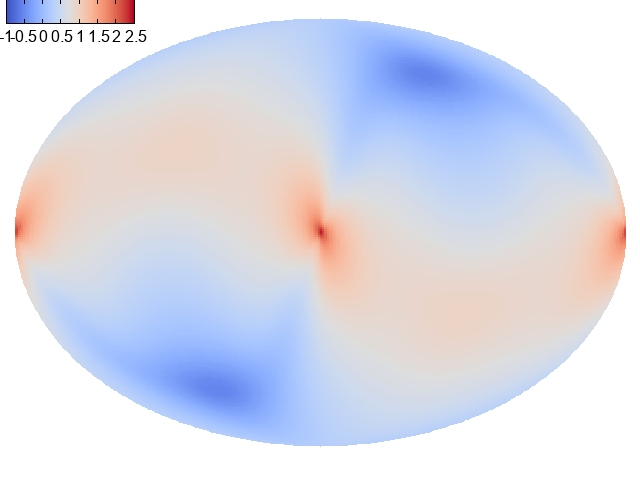}\label{fig:power-kibble-turok}
    }
    \hspace{1em}
    \subfloat[][Loop \#246, 10\% evaporation]{%
        \includegraphics[width=0.4\textwidth]{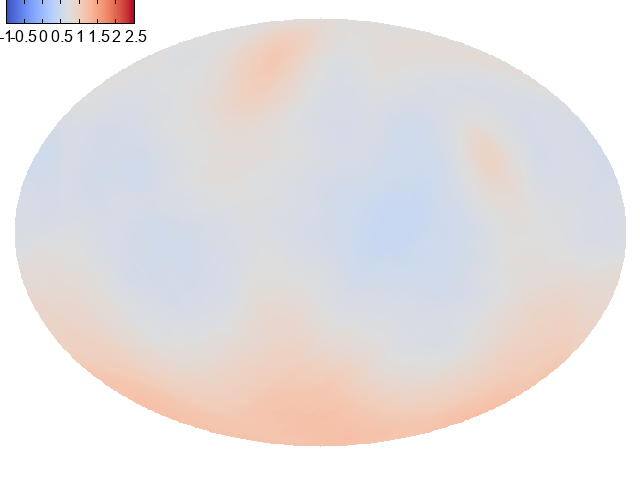}\label{fig:power-246-10}
    }

    \subfloat[][Loop \#246, 40\% evaporation]{%
        \includegraphics[width=0.4\textwidth]{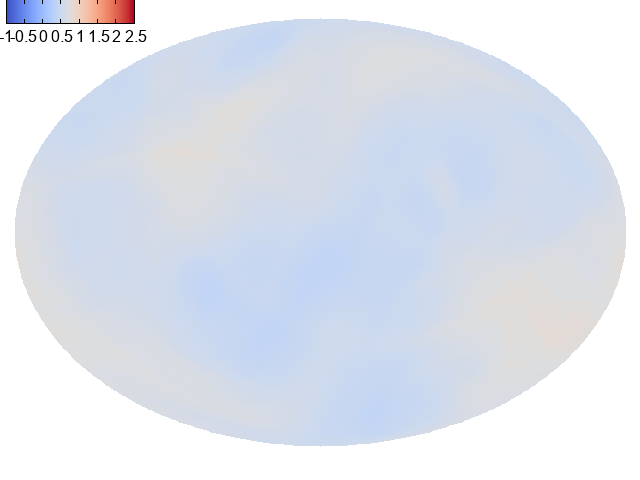}\label{fig:power-246-40}
    }
    \hspace{1em}
    \subfloat[][Loop \#246, 70\% evaporation]{%
        \includegraphics[width=0.4\textwidth]{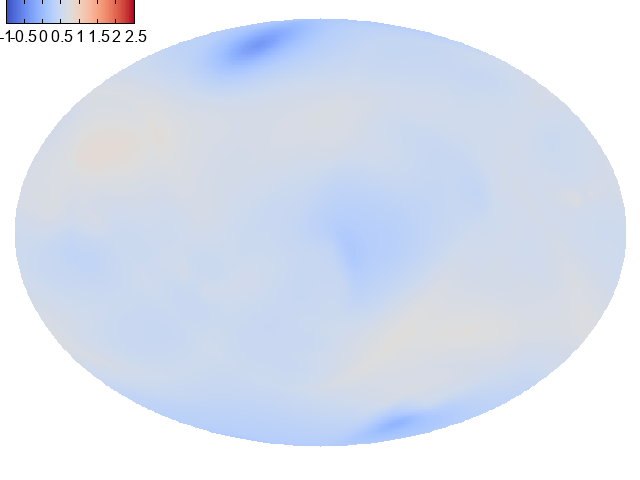}\label{fig:power-246-70}
    }
    \caption{Gravitational wave strength radiated to infinity in all directions for a canonical Kibble-Turok loop (a) compared to one of our corpus' loops at different stages of evaporation (b)--(d). We map the sphere at infinity to 2D via the Mollweide projection. The color bars show the base-10 logarithm of the intensity of the radiation. The Kibble-Turok cusp is far stronger than any localized feature in the corpus loop. The corpus loop instead emits GWs more uniformly in all directions.}\label{fig:cusp-comparison-individual}
\end{figure}

If we consider the radiation of gravitational waves onto the sphere at infinity by a loop, we can see the lack of strong cusps in loops from our corpus. We take as an example our Loop \#246, the most-segmented loop taken to $\chi=0.7$ that experiences no self-intersections. It has 429 segments in total and was produced at $\tau=357.6$.  We compare this loop, at various steps of $\chi$, to the canonical Kibble-Turok loop (which has two cusps and no kinks) without any backreaction.\footnote{See~\cite{Blanco-Pillado:2019nto} for details on how the canonical Kibble-Turok loop changes due to backreaction.} We chose the canonical Kibble-Turok loop as a ``maximally cuspy'' loop, i.e., one whose total gravitational power emission is dominated by its cusps. This can be seen in Fig.~\ref{fig:power-kibble-turok}, where the two concentrated red spots indicate the (antipodal) directions in which the Kibble-Turok loop's cusps beam the gravitational radiation.

By way of comparison, we show in the rest of Fig.~\ref{fig:cusp-comparison-individual} Loop \#246 at $\chi=0.1,0.4,0.7$. For these evaporation fractions, this loop has only two cusp candidate crossings. The lack of concentrated GW emission to infinity by this loop indicates that there are no cusps of analogous strength to the Kibble-Turok case.

In making this comparison, we should keep in mind the logarithmic (base-10) scale used in visualizing the results. The Kibble-Turok loop, at a region containing a cusp, has a maximum strength about ten times larger than the largest value in any region for the corpus loop at 10\% evaporation. The 40\% and 70\% evaporated loops have smaller values still. The $\Gamma$ of these loops are comparable: the Kibble-Turok has $\Gamma\approx 68$, and the 10\%-evaporated loop has $\Gamma\approx 81$ (decreasing to $\approx 51, 45$ at the higher evaporations). While the total powers are then fairly close, the corpus loop's power is much more smeared across the sphere; the Kibble-Turok loop sets the minimum and maximum of the scale seen here, with the corpus loop's range in power always narrower.

All in all, the average loop in a network starts out without cusps, and the cusps it acquires never become particularly strong and are further weakened over time. Thus, we conclude that cusps do not meaningfully contribute to the typical loop's gravitational-wave radiation. 

As we mentioned earlier, cusps are a source of an intense beam of gravitational radiation that could lead to burst-like events in our detectors \cite{Damour:2000wa,Damour:2001bk,Damour:2004kw}. However, our calculations here suggest that the amount of energy involved in these bursts could be substantially lower than previously estimated. Therefore, it is clear that one will have to re-evaluate the prospects of detection of gravitational wave bursts coming from these cusps in current as well as planned gravitational wave detectors \cite{Yonemaru:2020bmr,LIGOScientific:2021nrg,LISAConsortiumWaveformWorkingGroup:2023arg}. 

\subsection{The evolution of string smoothness}\label{ssec:kink-angles}

For our loops, we can get some sense of smoothing due to backreaction by studying how the angles between segments of our loops change over time. As we use a piecewise linear model for our loops, we do not distinguish between ``true'' kinks, which would also be discontinuities on a real loop, and ``discretization'' kinks, which arise when taking a piecewise approximation of a curve.

\begin{figure}
    \centering
    \includegraphics[scale=1.00]{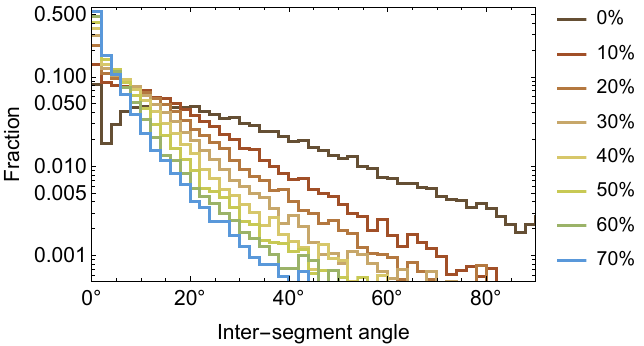}
    \caption{The distribution of angles between consecutive segments of the string's worldsheet functions, $A$ and $B$, for various evaporation stages. Bins are $2^\circ$ wide. The median angle at 0\% evaporation is $22.2^\circ$ and the distribution is spread over a range of moderate angles, with only a few very small angles. Very quickly, the majority of angles become very small and the distribution tends towards exponential; by 50\% evaporation, the median angle is $3.2^\circ$. While a few large kinks persist to high evaporation fractions, initially wiggly strings become smooth.}
    \label{fig:kink-changes}
\end{figure}

For our 70\% evaporated no-majors subpopulation, we find the angles between  consecutive $A$ segments and  consecutive $B$ segments across the loops' lifetimes, pool them all together by $\chi$, and examine how the distribution of angles changes with $\chi$. This is visualized in Fig.~\ref{fig:kink-changes}.

The initial distribution of angles is peaked around $14^\circ$, with a median of $22.2^\circ$, and extends up to moderately large values.\footnote{Note that the spike in the lowest bin for the 0\% evaporated curve is due to the segment-splitting procedure discussed in Sec.~\ref{ssec:kink-smoothing}: there are a number of zero-angle false kinks introduced as a result of this procedure which are quickly bent to non-zero angles by backreaction.} The effect of backreaction here is to reduce the typical angle between subsequent segments of the string worldsheet functions. A few large angles always persist---the largest angle at 70\%, for example, is $163^\circ$, corresponding to a kink in physical space of just over $80^\circ$---but the majority quickly drop to small angles; e.g., by about 12\% evaporation, half of all angles are less than $10^\circ$. This is consistent with the smoothing picture illustrated in Fig.~\ref{fig:loop-comparison}.

\subsection{The effects of kinks and small-scale structure on the power spectrum}\label{ssec:kinks}

There remains a distinction between the loops we study and loops in nature. At formation, our loops have structure on all scales, but there is a preferred scale coming from the initial conditions and a lower density of kinks at smaller scales.  Real loops have many more kinks, because the density of kinks does not scale in Nambu-Goto dynamics~\cite{PhysRevD.43.3173,Copeland:2009dk,Kawasaki:2010yi,Matsui:2016xnp}. The kink number begins to grow at the end of friction domination in the very early universe and is then limited by gravitational smoothing on long strings, but the range of scales is much larger than we can simulate. Thus extrapolating our results is important both for predicting the GWB and for understanding the rocket effect.

\begin{figure}
    \centering
    \includegraphics[scale=0.6]{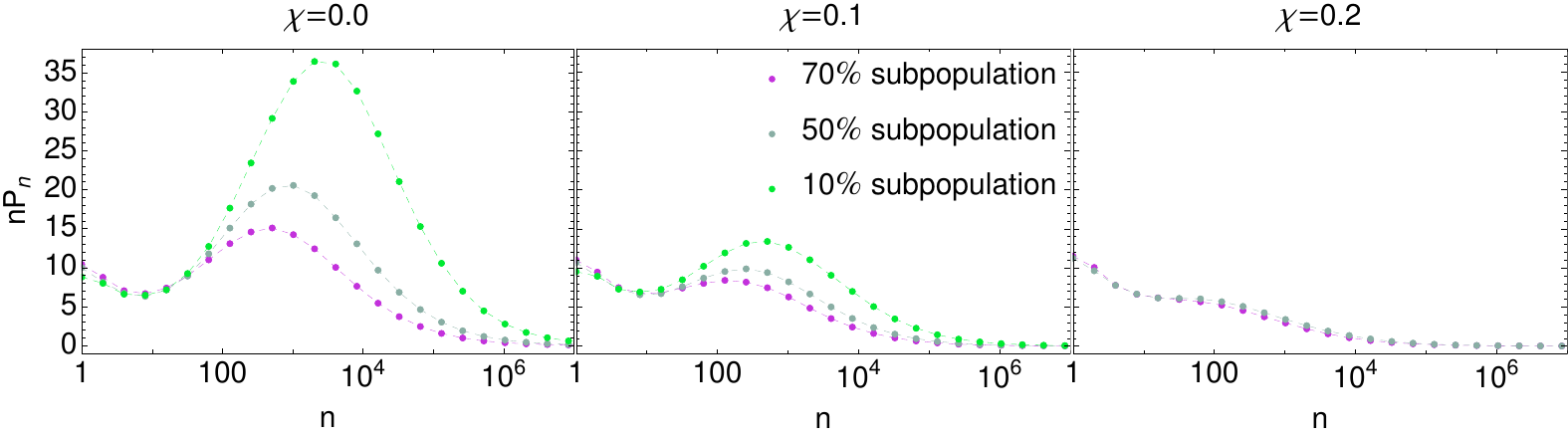}
    \caption{Loop spectra by \%-evaporated subpopulation for $\chi=0.0,0.1,0.2$. Though initially different due to differences in subpopulation segmentation, the spectra quickly converge as the loops are smoothed by backreaction. By the time we reach $\chi=0.2$, the 70\% and 50\%-evaporated subpopulations have effectively converged.}\label{fig:Pn-population-binned}
\end{figure}

From Fig.~\ref{fig:Pn-by-evaporation}, we can see that the most rapid change to the small-scale-structure bump in the power spectrum happens at low evaporation fractions. We focus now on understanding how the spectrum changes with segmentation for $\chi=0.0$ and $\chi=0.1$, making use of all loops in the no-majors subpopulation and partitioning them into the 70\%-evaporated, 50\%-evaporated, and 10\%-evaporated subpopulations as described in Table~\ref{tbl:evaporations}. As before, we can construct average spectra for each of these subpopulations at fixed $\chi$ and observe any trends with segmentation. This information is found in Fig.~\ref{fig:Pn-population-binned}.

For $\chi=0.0$, we see a clear trend in the bump's height and location to both increase as we go from the 70\% to the 50\% to the 10\% subpopulation, which is best understood by noting that this is an increase in the segment count. Then, the increase in location makes sense based on the previously-mentioned idea that significant structure at a scale of $L/n$ leads to a bump at $n$ in the power spectrum. The increase in height can be understood from the requirement that a loop with more structure radiate more power into gravitational waves---has a larger $\Gamma$---and so the larger bump is necessary to achieve a larger $\Gamma$, as the loops all look the same (and have the same $P_n$) at small $n$.

By $\chi=0.1$, the spectra for all three subpopulations have significantly reduced bump heights and downwards shifts in bump location. Proportionally, the 10\% subpopulation has experienced the most significant change, the 50\% the second-most, and the 70\% the last: the ratio of bump height at $\chi=0.0$ to bump height at $\chi=0.1$ is $2.74$ for 10\%, $2.09$ for 50\%, and $1.80$ for 70\%; the ratios of bump location at the same for the same are $6.25$, $3.36$, and $3.55$.\footnote{The 70\% subpopulation having a proportionally greater change in bump location can be explained by noting that the bump is, by $\chi=0.1$, ``turning over''; cf. the $\chi=0.2$ plot, where no local maximum can be associated to the bump.} 

By $\chi=0.2$ (noting that we must drop the 10\% subpopulation, which was evaporated only to $\chi=0.1$), the 70\% and 50\% spectra are effectively indistinguishable; agreement only improves for larger $\chi$. We conjecture that the 10\% subpopulation, if evolved to greater $\chi$, would converge to a similar final result. While there are significant differences in the loop $nP_n$ at low $\chi$ due to the difference in initial segmentation (kinkiness), as backreaction straightens out the strings and smooths the bump structure out, all loops spectra look the same at high evaporation fractions.

Having established the behavior of average spectra, when binned by segment count, we turn to looking at how the average spectra change with time of creation. For this particular investigation, we are not concerned with how the spectra change due to backreaction---if we can understand the dependence between $\tau$ and segment count, the results of Fig.~\ref{fig:Pn-population-binned} lets us predict how the spectra will change in $\chi$ at different times. Thus, it behooves us to use a data-set with a slightly larger range of $\tau$ than we have been.

\begin{figure}
    \centering
    \includegraphics[scale=1.0]{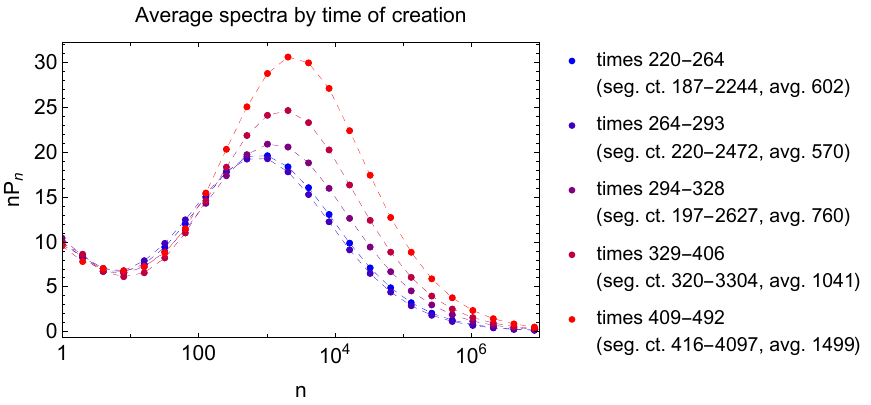}
    \caption{Initial spectra of the loops in an expanded corpus of 282 members in total, binned by conformal time of creation. Loops which form at later times tend to have a larger bump in their spectrum (at larger mode number $n$) as well as more segments.}\label{fig:Pn-time-binned-5}
\end{figure}

For this discussion only, we look at a corpus of 282 loops which includes the 198 loops in our main corpus, plus 84 additional loops formed at $215<\tau<250$ and $400<\tau<500$.\footnote{All of these loops were formed in the same symmetry-breaking simulation and are part of the same network; the $250<\tau<400$ set was chosen as a computationally tractable subpopulation, as discussed before.} These additional loops were selected to give greater than a factor of two range in the initial $\tau$s. We now sort our loops by conformal time of creation, then partition them as before. The result is shown in Fig.~\ref{fig:Pn-time-binned-5}.

Here we see that loops which form at later times have a higher bump in their spectra, and at a larger $n$. This binning also reveals that loops with more segments tend to form at later times, which is expected since the kink density on strings does not scale. As with the binning by total segment count, these differences are entirely in small-scale structure; the lowest-$n$ parts of the spectra are the same.

The results of this section allow us to extrapolate from our studies here to real, cosmological loops.  Real loops would have an even larger $\Gamma$, with a higher peak in $P_n$ at higher harmonic number $n$.  However, the difference will be short-lived.  The small structure causing the peak will be rapidly eliminated by backreaction, and at some quite low evaporation fraction, the backreaction would join on to what we observed in our population.  Thus the overall picture of evolution is correct, but the process starts at an earlier stage with a very short period of more intense, higher frequency radiation.   As we will discuss further in the companion paper~\cite{sgwb-paper}, this very early radiation does not affect the GWB.

\section{Conclusion}\label{sec:conc}

Gravitational backreaction smooths out the small-scale structure of strings while leaving the overall shape of the loop mostly intact. This effect is visible in both the loop's gravitational-wave power spectrum and the physical appearance of the loop. Despite this smoothing, cusps which form on initially-cuspless loops are very weak due to having a small length of string involved in them. The loop power spectrum does not appear cusp-like even when compared to an upper bound on the contribution from cusps.

Backreaction fairly regularly leads to self-intersections, but these typically lead to the emission of small ``looplets''; only in very rare cases does a mother loop fragment into comparably-sized daughter loops. Still, looplet emission is sufficient to cause a large recoil velocity on the typical loop, above about $0.001c$, which may inhibit loops from clustering in galaxies. A fuller study of this effect will be done in a subsequent paper.

The careful extrapolation of these results to real loops requires further study, which will be pursued in a companion paper~\cite{sgwb-paper} and subsequent work. This is due to differences in the small-scale structure to be expected on real loops from what we have on our loops. However, given what we know about the scale at which gravitational backreaction acts, as well as the distribution of loops which are a certain percentage evaporated, we expect the results reported on here to be fairly accurate to reality.

\section{Acknowledgments}

We thank Bruce Boghosian, Ali Masoumi, Alex Vilenkin, and Tanmay Vachaspati for useful conversations.  K. D. O. was supported in part by NSF Grant Nos.\ 2111738 and 2412818. J. J. B.-P. is supported by the PID2021-123703NB-C21 grant funded by MCIN/ AEI /10.13039/501100011033/ and by ERDF; ``A way of making Europe'', the Basque Government grant (IT-1628-22), and the Basque Foundation for Science (IKERBASQUE). The authors acknowledge the Tufts University High Performance Computing Cluster (\url{https://it.tufts.edu/high-performance-computing}), which was utilized for the research reported in this paper.

\appendix

\section{Calculation of the cusp power spectrum coefficient}\label{app:cusp-calculation}

In this appendix, we derive the formula used to find the constant prefactor in $P_n = Qn^{-4/3}$ for a cusp-like spectrum.

For each cusp, we start from Eq.~(A29) of~\cite{Blanco-Pillado:2017oxo}, which we repeat here:
\begin{align}
    \frac{dP}{d\omega d\Omega}&= \frac{2G \mu^2 \omega^2\theta^8}{9 \pi^2L}\frac{\sin^4\phi_+\sin^4\phi_-}{\alpha_+^2\alpha_-^2} \bigg[\left(K^2_{1/3}(\xi_+) +K^2_{2/3}(\xi_+)\right)\left(K^2_{1/3}(\xi_-) +K^2_{2/3}(\xi_-)\right)\\
&\qquad\qquad\qquad\qquad+4\sign(\sin\phi_+\sin\phi_-)K_{1/3}(\xi_+)K_{2/3}(\xi_+)K_{1/3}(\xi_-)K_{2/3}(\xi_-)\bigg]\,.\nonumber
\end{align}
The parameters and functions have the following meanings: $L$ and $G\mu$ retain their meanings as length and coupling to gravity; $\omega$ indicates angular frequency; $\theta$ is the angle between the cusp direction and the observer direction; the $\alpha_\pm$ and $\phi_\pm$ describe the magnitude and angle relative to the observer direction of $\vec X''_\pm$, where in the language used in this paper $2\vec X''_+\equiv \vec B''$ and $2\vec X''_-\equiv \vec A''$; we define $\xi_\pm = \omega\theta^3\left|\sin^2\phi_\pm\right|/(6\alpha_\pm)$; and $K$ is the modified Bessel function. The $\alpha_\pm$ and $\phi_\pm$ describe the crossing of the tangent vectors on the unit sphere and thus the strength of the cusp.

Our goal is to find $P_n$, a function of the mode number, which can be understood as $(dP/dn)/(G\mu^2)$ (in that $\Gamma = \sum_n P_n$). To obtain $P_n$, we must make the substitution $\omega = 4\pi n/L$ and integrate across the sphere for a fixed mode number. First making the change to $n$, we get
\begin{align}
    \frac{dP_n}{d\Omega}&= \frac{128\pi G \mu^2 n^2\theta^8}{9 L^4}\frac{\sin^4\phi_+\sin^4\phi_-}{\alpha_+^2\alpha_-^2} \bigg[\left(K^2_{1/3}(\xi_+) +K^2_{2/3}(\xi_+)\right)\left(K^2_{1/3}(\xi_-) +K^2_{2/3}(\xi_-)\right)\label{eqn:dPndOm}\\
&\qquad\qquad\qquad\qquad+4\sign(\sin\phi_+\sin\phi_-)K_{1/3}(\xi_+)K_{2/3}(\xi_+)K_{1/3}(\xi_-)K_{2/3}(\xi_-)\bigg]\,.\nonumber
\end{align}
The cusp strength is fixed, so $\alpha_\pm$ are constants. The entire term in brackets depends on both $\theta$ and $\phi$ (the azimuthal coordinate).

Some definitions are useful for simplifying expressions below. First define a new variable $\eta=4\pi n\theta^3/L$, and then
\be
    f_\pm = \frac{\left|\sin^3\phi_\pm\right|}{6\alpha_\pm}\,,
\ee
so $\xi_\pm = \eta f_\pm$. From here, define
\be
    a = \frac{f_+}{f_-} = \frac{\alpha_-}{\alpha_+}\left|\frac{\sin\theta_+}{\sin\theta_-}\right|^3\,,\qquad z=\sqrt{f_+f_-}\eta\,.
\ee
With this setup, we then rewrite Eq.~(\ref{eqn:dPndOm}) as
\be
    \frac{dP_n}{d\Omega}= \frac{128\pi n^2\theta^8}{9 L^4}\frac{\sin^4\phi_+\sin^4\phi_-}{\alpha_+^2\alpha_-^2} h_\pm(a,z)\,,
\ee
where
\begin{align}
    h_\pm(a,z) = &\left(K^2_{1/3} (a^{-1/2}z)+K^2_{2/3}(a^{-1/2}z)\right)\left(K^2_{1/3} (a^{1/2}z)+K^2_{2/3} (a^{1/2}z)\right)\\
    &\pm4 K_{1/3} (a^{-1/2}z)  K_{2/3} (a^{-1/2}z) K_{1/3} (a^{1/2} z) K_{2/3} (a^{1/2} z)\,,\nonumber
\end{align}
which is invariant under $a\rightarrow 1/a$. Additionally, define
\be
    \mathcal{H}^{(3)}_\pm(a) = \int^\infty_0 z^{7/3}h_\pm(a,z)\,dz\,.
\ee
Now, we proceed with
\be
    P_n = \int^{2\pi}_0\int^{\theta_1}_0 \theta \frac{dP_n}{d\Omega}\,d\theta\,d\phi\,,
\ee
where $\theta_1$ is some upper limit which sets the opening angle of the cone within which we calculate the power from a cusp. The cusp power in a particular mode falls off exponentially outside of a fairly narrow angle~\cite{Damour:2001bk}, and so above some $\theta_c\approx n^{-1/3}$, increasing $\theta_1$ makes no difference in the overall result. Since $n\geq 1$, $\theta_c\lesssim 1$. Thus, all choices for $\theta_1\gtrsim 1$ give equivalent final values. With this justification, we take $\theta_1\rightarrow\infty$, as the asymptotic forms of several functions involved are easier to work with and more numerically robust. 

To do this integral more easily, we change variables to $z$, with $dz = 12\pi n\sqrt{f_+f_-}\theta^2d\theta$, and substitute $\sqrt{f_+f_-}\rightarrow (1/6)\sqrt{\sin^3\phi_+\sin^3\phi_-/(\alpha_+\alpha_-)}$. Then, after some simplification,
\be
    P_n = \int^{2\pi}_0 \frac{16(3/2)^{1/3}\mathcal{H}^{(3)}_{\sign(\sin\phi_+\sin\phi_-)}(a)}{\pi^{7/3}n^{4/3}L^{2/3}(\alpha_+\alpha_-)^{1/3}|\sin\phi_+\sin\phi_-|}\,d\phi\,.
\ee
Thus,
\be
    Q = \frac{32(3/2)^{1/3}}{\pi^{7/3}L^{2/3}}\int^\pi_0\frac{\mathcal{H}^{(3)}_{\sign(\sin\phi_+\sin\phi_-)}(a)}{(\alpha_+\alpha_-)^{1/3}|\sin\phi_+\sin\phi_-|}\,d\phi\,,
\ee
where we have used the symmetry in the integrand to halve the range of integration of $\phi$.

Finally, we sum over all possible cusps, i.e., all crossings on the unit sphere when we draw lines between successive values of the $A'$ and $B'$, to give an overall $Q$ for each loop.

\bibliography{paper}

\end{document}